\documentclass[
 amsmath,amssymb,
 aps,
reprint,prl,twocolumn,twoside,aps,superscriptaddress]{revtex4-2}
\usepackage{soul} 
\usepackage[normalem]{ulem}
\usepackage{graphicx}
\usepackage{dcolumn}
\usepackage{bm}
\usepackage{physics}
\usepackage{xcolor}

\usepackage{hyperref}
\usepackage{orcidlink}
\usepackage{comment}

\begin{document}

\preprint{APS/123-QED}

\title{Observation of resonant monopole-dipole energy transfer between Rydberg atoms and polar molecules}

\author{J. Zou\orcidlink{0000-0002-0629-9520}}
\affiliation{Department of Physics and Astronomy, University College London, Gower Street, London WC1E 6BT, United Kingdom}
\author{R. R. W. Wang\orcidlink{0000-0002-1069-9746}}
\email{reuben.wang@cfa.harvard.edu}
\affiliation{ITAMP, Center for Astrophysics $|$ Harvard \& Smithsonian Cambridge, Massachusetts 02138, USA}
\affiliation{ Department of Physics, Harvard University, Cambridge, Massachusetts 02138, USA }
\author{R. Gonz\'alez-F\'erez\orcidlink{0000-0002-8871-116X}}
\affiliation{Instituto Carlos I de F\'{i}sica Te\'{o}rica y Computacional, and Departamento de F\'{i}sica At\'{o}mica, Molecular y Nuclear, Universidad de Granada, 18071 Granada, Spain}
\author{H.~R.~Sadeghpour\orcidlink{0000-0001-5707-8675}}
\affiliation{ITAMP, Center for Astrophysics $|$ Harvard \& Smithsonian Cambridge, Massachusetts 02138, USA}
\author{S. D. Hogan\orcidlink{0000-0002-7720-3979}}
\email{s.hogan@ucl.ac.uk}
\affiliation{Department of Physics and Astronomy, University College London, Gower Street, London WC1E 6BT, United Kingdom}

\date{\today}

\begin{abstract}
Resonant energy transfer (RET), between the equal parity 1s65s\,$^3\mathrm{S}_1$ and 1s66s\,$^3\mathrm{S}_1$ Rydberg levels in helium has been observed in low-temperature ($\sim80$~mK) collisions with ammonia molecules which undergo inversion transitions in their X$\,^1$A$_1$ ground electronic state. This hybrid Rydberg-atom--polar-molecule RET represents a monopole-dipole energy exchange reaction that necessarily requires spatial overlap of the Rydberg-electron and molecular wavefunctions. Calculations, that account explicitly for the charge-dipole interaction between the Rydberg electron and the molecule, provide a quantitative explanation of the observations.
Total parity is conserved in the reaction through the mixing of collisional angular momentum in the atom-molecule complex. This work opens opportunities to expand the toolbox for quantum science with charge-dipole--mediated energy exchange in hybrid neutral-atom--polar-molecule platforms.
\end{abstract}

\maketitle

Resonant energy transfer, the phenomenon by which an excitation is exchanged between energetically resonant quantum states, is of widespread interest in physics, chemistry and biology~\cite{jones2019resonance}. 
It was first discussed by F\"orster~\cite{forster1946energy} in the context of energy migration in dye polymers.
Since pioneering experiments of Cario and Frank~\cite{cario1922uber}, RET has been utilized to explain photosynthesis in light harvesting complexes~\cite{sener2011forster, mirkovic2017light}, for imaging at single-molecule resolution~\cite{ha2024flourescence}, to tune molecular interactions and suppress chemical reactions~\cite{li2021tuning, lassabliere2022model}, and to enrich the toolbox for quantum information processing~\cite{carroll25observation,guttridge23observation, picard2025entanglement}. 

\begin{figure}[ht!]
\includegraphics[width=0.92\columnwidth]{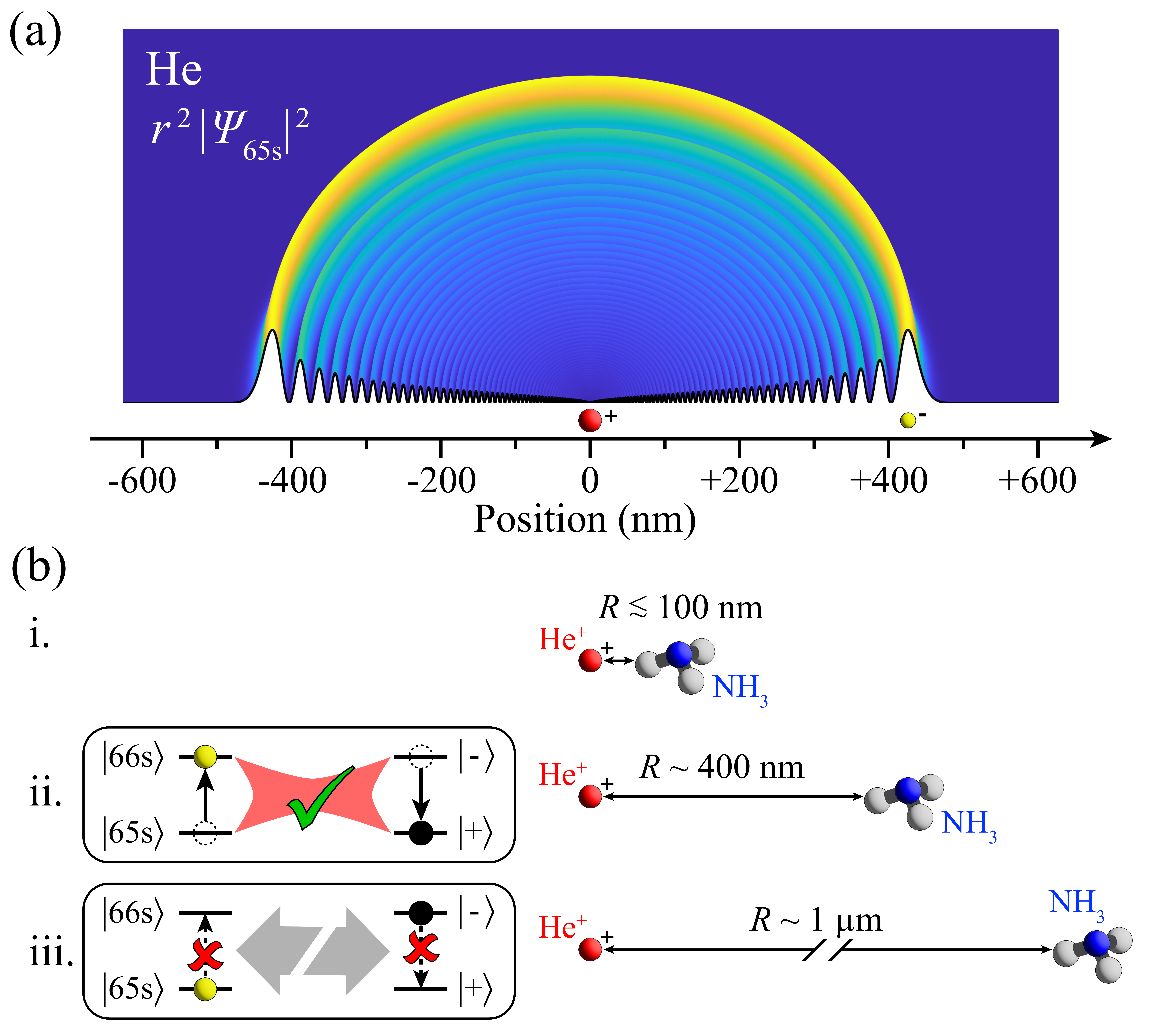}
\caption{(a) Electron charge distribution in the $|65\mathrm{s}\rangle$ orbital in He. (b) Distance scales associated with the interaction of a Rydberg He atom and an NH$_3$ molecule at which the interaction of the molecule with the (i) He$^+$ ion core, (ii) Rydberg electron, and (iii) composite neutral atom dominate. The resonant monopole-dipole energy transfer reported here is forbidden at large distances (iii-left), but can occur at intermediate distances because of the charge-dipole interaction with the Rydberg electron (ii-left).\label{fig1}}
\end{figure}

FRET (F\"orster RET), which relies on the interaction between electric dipole transition moments~\cite{perrin1927fluorescence, kallmann1929on, perrin32theory,forster1946energy}, has been studied in cold dilute gases with Rydberg-Rydberg~\cite{vanDitzhuijzen2008spatially, ravets2014coherent, deLeseleuc2017optical}, Rydberg--polar-molecule~\cite{zhelyazkova2017electrically,jarisch2018state,zou2022probing,patsch2022rydberg,zhu2025probing}, and molecule--molecule~\cite{yan2013observation, christakis2023probing} interactions. Beyond dipolar interactions, RET can also be associated with higher-order multipole transition moments, as evident, e.g., from observations of RET due to dipole-quadrupole couplings~\cite{gallagher80resonant, deiglmayr2014observation, maineult2016dipole}. There is currently growing interest in RET between Rydberg atoms and polar ground-state molecules since the hybridization of these two platforms for quantum information processing, through resonant exchange interactions, would allow, e.g., rotational states in the molecules to be exploited as long-coherence-time quantum memories with initialization or readout through the atoms~\cite{wang22enriching, zhang22quantum}.

Here, we report the observation of a RET process that involves a monopole transition in a Rydberg atom, and an electric dipole transition in a polar molecule. 
The term monopole here refers to an electronic transition between equal parity Rydberg states both of which have zero electron orbital angular momentum, and not to the order of expansion in the charge distribution.
Specifically, this involves the $|65\mathrm{s}\rangle\leftrightarrow |66\mathrm{s}\rangle$ transition between equal parity triplet Rydberg states in He ($|n\mathrm{s}\rangle\equiv\,|1\mathrm{s}n\mathrm{s}\,^3\mathrm{S}_1\rangle$), and the opposite-parity states of the inversion doublet, $|\pm\rangle$, in ammonia (NH$_3$) [X$\,^1$A$_1(J=K=1)$ where $J$ and $K$ are the rotational quantum numbers].
The single-photon transition between these Rydberg states is forbidden by both electric dipole and quadrupole  selection rules. Consequently, RET between these systems must be accompanied by a change in collisional angular momentum to conserve total parity. From the perspective of the composite system comprising the polar molecule located inside the Rydberg orbital, the results presented here are analogous to electric-monopole transitions associated with internal conversion in atomic nuclei~\cite{church56a,zerguine08a}. They represent, to our knowledge, the first observation of RET in a cold neutral gas that does not rely on interactions at large interparticle separations.

When a Rydberg atom and a neutral molecule are far apart, as in Fig.~\ref{fig1}(b-iii), they interact \emph{via} far-field electrostatic interactions, either by first order induced or permanent dipole moments, or second order dispersion interactions, leading to level shifts, excitation blockade, and, under appropriate conditions, RET~\cite{deLeseleuc2017optical,zhu2025probing,christakis2023probing}.
When a polar molecule is in the vicinity of the Rydberg atom, near the outer lobe of the Rydberg-electron charge distribution [compare Fig.~\ref{fig1}(a) and (b-ii)], charge-dipole interactions dominate.
Within the Rydberg electron orbit, these interactions must be treated in the near-field where the Rydberg atom is no longer a point charge distribution~\cite{Kuznetsova2016rydberg,rittenhouse2010ultracold, Schlagmuller16ultracold, Geppert2021diffusive}. 
Finally, when the molecule is further inside the Rydberg electron charge distribution [Fig.~\ref{fig1}(b-i)], it interacts predominantly with the ion core, often leading to barrier-less exothermic ion-molecule reactions~\cite{zhelyazkova21multipole}. This region we simply refer to here as short distances.

We show here that the interaction of a Rydberg electron with the dipole moment of a polar molecule can mediate resonant monopole-dipole energy transfer. The theory that describes this RET also explains how off-resonant energy transfer between the same pair of molecular states, and the $|65\mathrm{s}\rangle\leftrightarrow|64\mathrm{s}\rangle$ transition in He is suppressed. 
This work demonstrates a new probe of atom-molecule interactions at mesoscopic scales (on the order of hundreds of nanometers), expanding the toolbox for quantum information processing in hybrid Rydberg-atom--molecule platforms.

The experiments were performed in an intrabeam collision apparatus with pulsed supersonic beams contained pure He, or a mixture of NH$_3$ and He (1:63 by pressure)~\cite{gawlas2019rydberg}. The He atoms were prepared in the metastable 1s2s\,$^3\mathrm{S}_1$ level using a DC electric discharge~\cite{halfmann2000source}. After collimation and the removal of charge particles, the atoms were excited to $|n\mathrm{s}\rangle$ Rydberg states ($64 \leq n \leq 67$) using the resonance-enhanced two-color two-photon 1s2s\,$^3\mathrm{S}_1\rightarrow$1s3p\,$^3\mathrm{P}_2\rightarrow$1s$n$s\,$^3\mathrm{S}_1$ scheme~\cite{hogan2018laser}. With the NH$_3$ in the beam, the molecule number density was $(1.5\pm0.4)\times10^{10}$~cm$^{-3}$ at the Rydberg state photoexcitation position, and the mean center-of-mass collision speed between the atoms and the molecules was $19.3\pm2.6$~m/s~\cite{zou2022probing}. Consequently, the relative translational temperature at which atom-molecule collisions occurred was $E_{\mathrm{kin}}/k_{\mathrm{B}}\sim80$~mK ($E_{\mathrm{kin}}/h\sim1.6$~GHz or $E_{\mathrm{kin}}/hc\sim0.055$~cm$^{-1}$), with $E_{\mathrm{kin}}$ the center of mass collision energy.

Following photoexcitation, the cold Rydberg atoms and the molecules evolved in nominally zero electric field for up to $12~\mu$s. Then, a slowly-rising electric field pulse was applied to ionize the Rydberg atoms. In this field, atoms in states with higher (lower) values of $n$ ionized at earlier (later) times. Consequently, the arrival time of the ionized electrons at a microchannel plate detector could be correlated with their ionization field, and the value of $n$ of the states populated. Stray electric fields in the interaction region of the apparatus were canceled to $<20$~mV/cm by microwave spectroscopy of the single-color two-photon $|65\mathrm{s}\rangle\rightarrow|63\mathrm{s}\rangle$ transition (see Appendix~\ref{app:expt}).

\begin{figure}
\includegraphics[width=0.70\columnwidth]{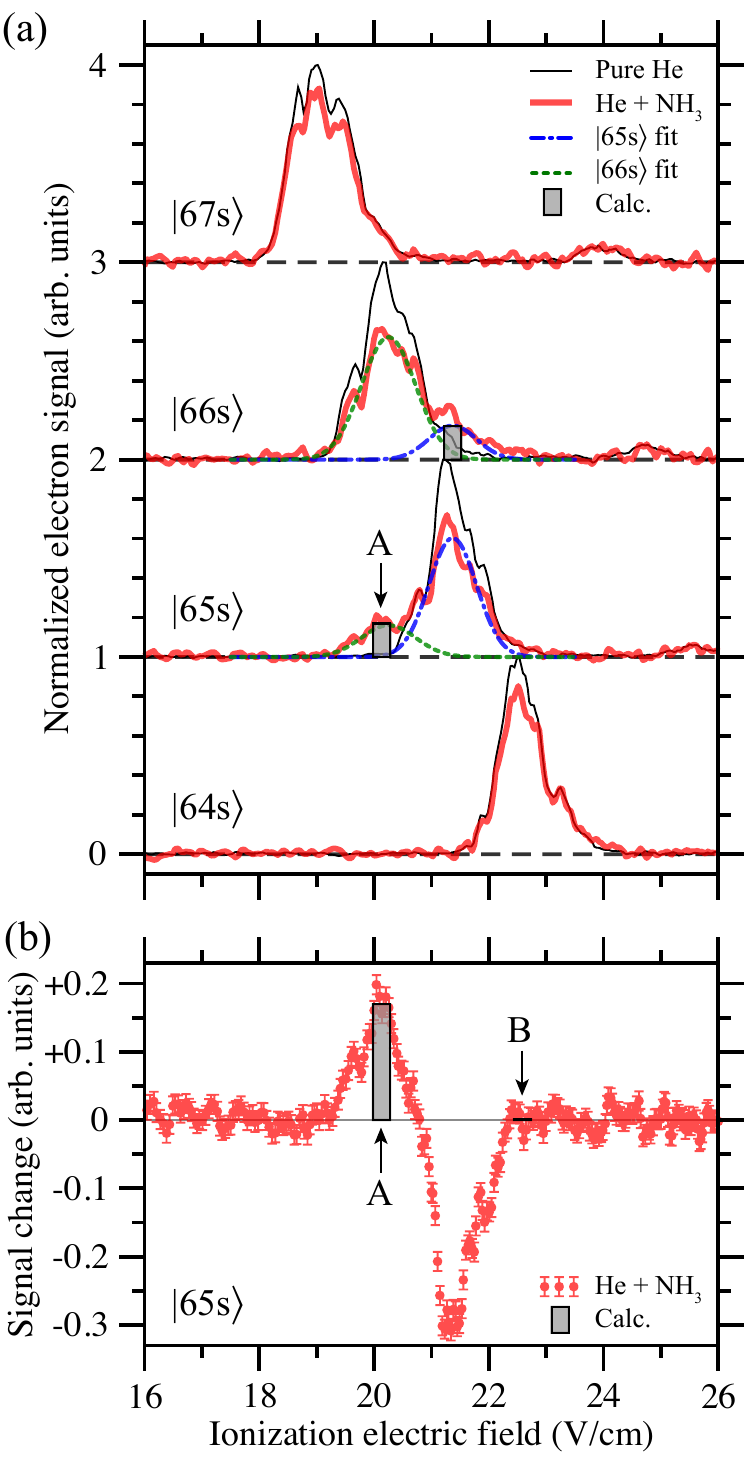}
\caption{(a) Electron signals recorded following excitation of He atoms to selected $|n\mathrm{s}\rangle$ Rydberg states in the absence (thin black curves) and presence (thick red curves) of NH$_3$. The horizontal axis indicates the electric field in which the Rydberg atoms ionized. The calculated population transfer for atoms initially prepared in the $|65\mathrm{s}\rangle$ and $|66\mathrm{s}\rangle$ states in the presence of NH$_3$ 
[$\mathbb{P}\left( |{ 65\mathrm{s},- }\rangle \rightarrow |{ 66\mathrm{s},+ }\rangle \right) \approx 17\%$] 
is indicated by the vertical bars. (b) Difference between the electron signals recorded without and with NH$_3$ present for atoms prepared in the $|65\mathrm{s}\rangle$ state, with the calculated resonant (off-resonant) population transfer to the $|66\mathrm{s}\rangle$ ($|64\mathrm{s}\rangle$) state labeled A (B)
[$\mathbb{P}\left( |{ 65\mathrm{s},+ }\rangle \rightarrow |{ 64\mathrm{s},- }\rangle \right) \approx 0.12\%$]. Note: The uncertainty in the calculations is not shown but is comparable to the uncertainty on the experimental data points.}\label{fig2}
\end{figure}

\begin{figure}
\includegraphics[width=0.95\columnwidth]{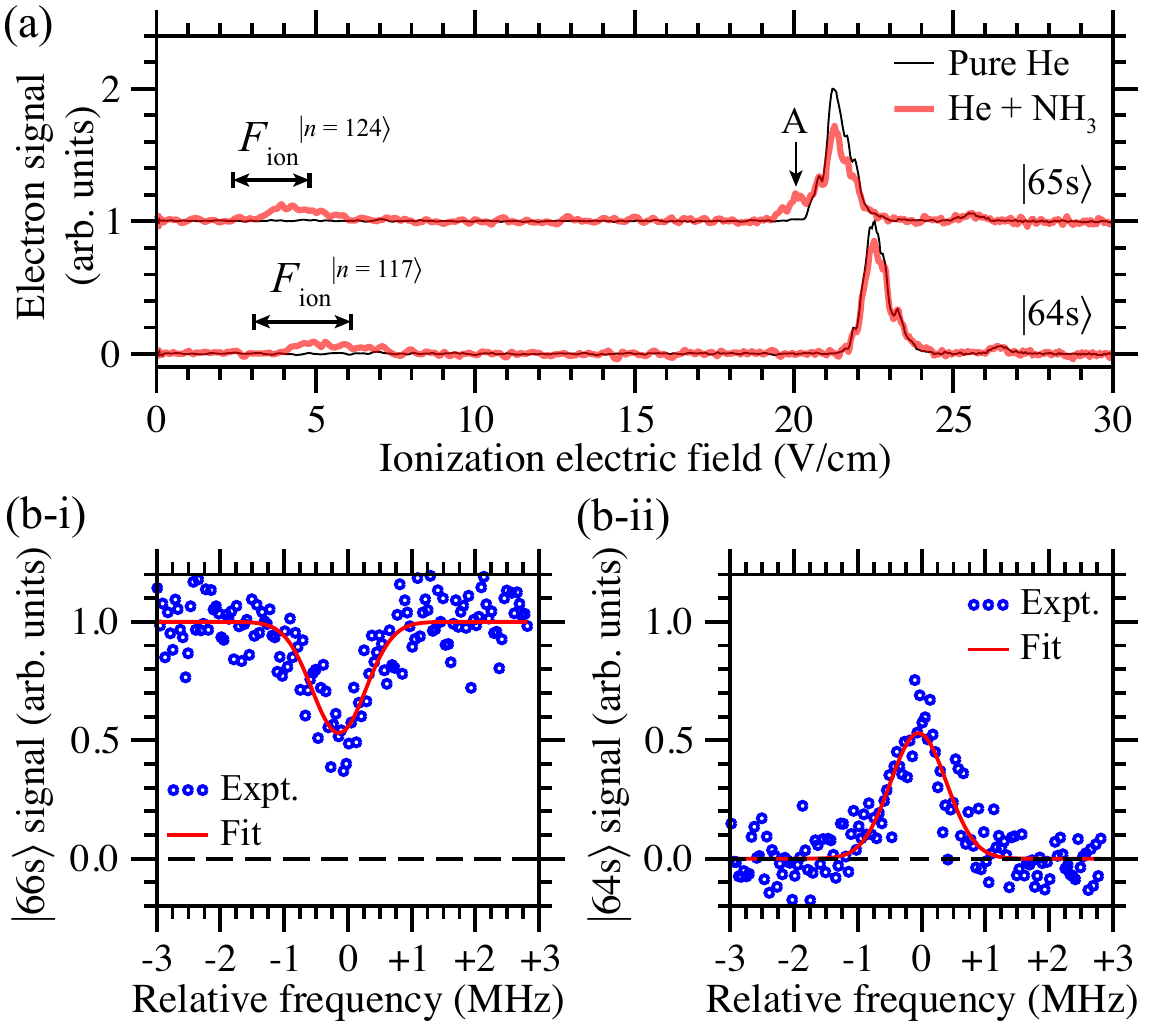}
\caption{(a) Broad view of the ionization signals in Fig.~\ref{fig2} for atoms prepared in the $|64\mathrm{s}\rangle$ and $|65\mathrm{s}\rangle$ states with (red) and without (black) NH$_3$. $F_{\mathrm{ion}}^{|n\rangle}$ denotes the range of ionization electric fields of the high-$n$ hydrogenic Rydberg states populated by electric-dipole-allowed rotational energy ($\sim572$~GHz) transfer from the NH$_3$. (b) Microwave spectra of the single-color two-photon $|66\rm{s}\rangle\rightarrow|64\rm{s}\rangle$ transition in atoms that underwent RET from the $|65\mathrm{s}\rangle$ state following collisions with NH$_3$. The microwave frequency is displayed with respect to the field-free two-photon transition frequency of 24.298\,146~GHz.}\label{fig3}
\end{figure}

\textit{Resonant monopole-dipole energy transfer} -- {To probe resonant monopole-dipole energy transfer between the molecules and atoms, data were recorded with He atoms prepared in selected $|n\mathrm{s}\rangle$ states. In the absence of NH$_3$ this resulted in the set of separated electric field ionization features with maxima at fields from $\sim22.5$ -- 19~V/cm indicated by the thin continuous black curves in Fig.~\ref{fig2}(a). With the molecules present (thick red curves), the normalized ionization profiles obtained following preparation of the $|64\mathrm{s}\rangle$ and $|67\mathrm{s}\rangle$ states maintain a similar structure, but some depletion is observed. This is a result of electric-dipole-allowed rotational energy transfer to states with values of $n>100$~\cite{smith78a}, which can be seen in Fig.~\ref{fig3}(a) where a broader range of ionization fields are spanned. However, for the $|65\mathrm{s}\rangle$ and $|66\mathrm{s}\rangle$ states the situation is different.} 
{For atoms in the $|65\mathrm{s}\rangle$ state in Fig.~\ref{fig2}(a), an appreciable electron signal is apparent in the region labelled A. This results from population transfer to the $|66\mathrm{s}\rangle$ state. A fit of the sum of two Gaussian functions (dashed green and dash-dotted blue curves) constrained by the fits to the pure-He reference data, yields a  transfer rate of $17\pm5\%$ from the $|65\mathrm{s}\rangle$ state to the $|66\mathrm{s}\rangle$ state which are separated by $23.735$~GHz. A similar behavior is seen upon preparation of atoms in the $|66\mathrm{s}\rangle$ state, with $16\pm5\%$ population transfer to the $|65\mathrm{s}\rangle$ state. 

The data in Fig.~\ref{fig2} demonstrate a resonant transfer of energy involving the $|-\rangle\leftrightarrow|+\rangle$ electric dipole allowed inversion transition in NH$_3$ at $23.695$~GHz~\cite{simmons48structure} (these states are equally populated in the supersonic beam), and the $|65\mathrm{s}\rangle \leftrightarrow |66\mathrm{s}\rangle$ transition in He. 
This seemingly violates parity conservation because it involves a pair of equal parity Rydberg levels, but molecular states with opposite parity.
Nevertheless, we confirm that the $|66\mathrm{s}\rangle$ state is indeed selectively populated in this process by high-resolution microwave spectroscopy of atoms that underwent energy transfer. This was implemented by applying a 2-$\mu$s-duration microwave pulse to probe the atoms using the single-color two-photon $|66\mathrm{s}\rangle\rightarrow|64\mathrm{s}\rangle$ transition at $2\times24.298\,146$~GHz, after an interaction time of $5~\mu$s with the molecules, while monitoring the $|66\mathrm{s}\rangle$ and $|64\mathrm{s}\rangle$ electron signals. The resulting spectra are shown in Fig.~\ref{fig3}(b). Gaussian functions fit to these data (continuous red curves) indicate that the measured transition frequency of $24.298\,050(41)$~GHz is $-96\pm41$~kHz below the field-free transition frequency. We attribute this shift to a residual stray electric field of $16.5\pm4.5$~mV/cm in the interaction region of the apparatus.}
This stray field induces a negligible ($<0.03\%$) admixture of $|n\mathrm{p}\rangle$ character into the $|n\mathrm{s}\rangle$ states, and is therefore not primarily responsible for the observed energy transfer (see Appendix~\ref{app:expt}). While all previous observations of RET have been interpreted through interactions occurring between the particles in the far-field, here, the resonant monopole-dipole (Rydberg-atom--polar-molecule) energy transfer is mediated by near-field interactions of the Rydberg electron with the molecular dipole.

\textit{Charge-dipole mediated RET} -- The anisotropic charge-dipole interaction potential has the form  \cite{Kuznetsova2016rydberg}:
\begin{align} \label{eq:charge_dipole}
    V_{\rm CD}(\boldsymbol{r}, \boldsymbol{R})
    &=
    -
    \boldsymbol{\mu}_{\rm NH_3}
    \cdot 
    \boldsymbol{{\cal E}}_{\rm Ryd}(\boldsymbol{r}, \boldsymbol{R}).
\end{align}
where $\boldsymbol{R}$ is the coordinate between the molecule and the He$^+$ ion core, 
\begin{align}
    \boldsymbol{{\cal E}}_{\rm Ryd}(\boldsymbol{r}, \boldsymbol{R})
    &=
    \frac{ e }{ 4 \pi \epsilon_0 }
    \left(
    \frac{ \boldsymbol{R} }{ R^3 }
    -
    \frac{ \boldsymbol{R} - \boldsymbol{r} }{ \abs{ \boldsymbol{R} - \boldsymbol{r} }^3 }
    \right)
\end{align}
is the electric field generated by the He$^+$ ion and the Rydberg electron at position $\boldsymbol{r}$, and $\boldsymbol{\mu}_{\rm NH_3}$ is the electric dipole operator coupling opposite parity states in NH$_3$.  
The charge-dipole interaction scales as $1/R^2$ and is in fact long-ranged. However, as shown below, it becomes important at near-field distances ($\lesssim400$~nm in the case considered here), when the molecule is located inside the Rydberg-electron charge distribution. In this situation, for the triplet $\ell=0$ Rydberg states in He with an energy splitting which is closest to resonance with the NH$_3$ inversion interval, this electron-molecule interaction is dominated by the monopole term. This term is non-zero because the corresponding integral only extends over a restricted range, i.e., $0<r<R$ (see Appendix~\ref{app:monopole_dipole} and the Supplemental Material~\cite{SI}.

\begin{figure}[t]
    \centering
    \includegraphics[width=\columnwidth]{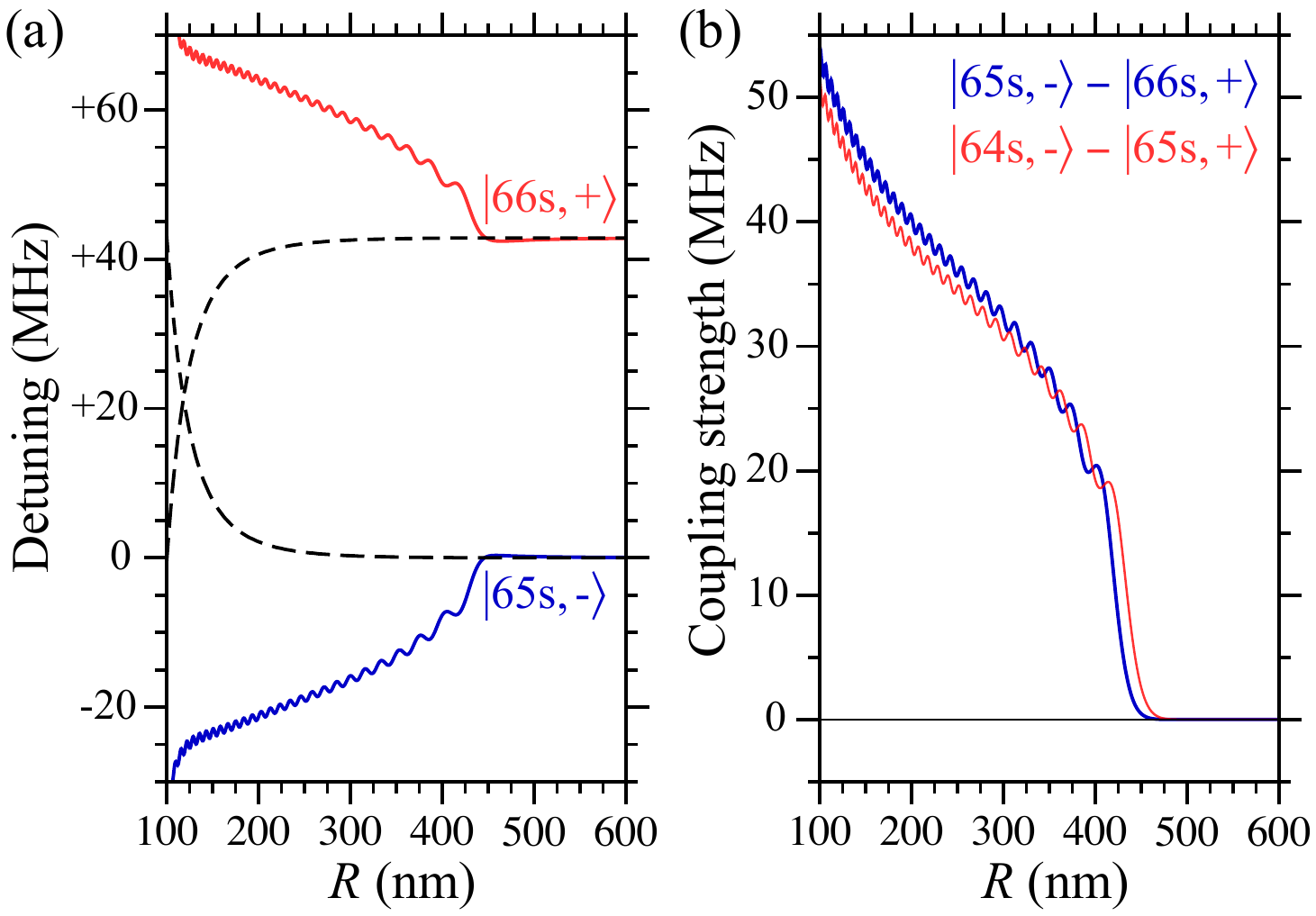}
    \caption{(a) Adiabatic (continuous curves) and diabatic (dashed curves) atom-molecule interaction potentials as a function of internuclear distance $R$. The curves corresponding to the $\ket{ 65\mathrm{s}, - }$ and $\ket{ 66\mathrm{s}, + }$ asymptotic eigenstates are indicated.
    The hybridization of the opposite parity molecular states becomes appreciable only near the outer turning point of the Rydberg electron wave function ($R\sim400$~nm) because of the enhanced charge-dipole interaction.
    (b) The coupling matrix element between the $\ket{ 65\mathrm{s}, - }$ and $\ket{ 66\mathrm{s}, + }$ states (continuous dark blue curve), and $\ket{ 64\mathrm{s}, - }$ and $\ket{ 65\mathrm{s}, + }$ states (continuous light red curve) as a function of $R$.}
    \label{fig:theory_figure}
\end{figure}

The adiabatic potential curves (adiabats) that connect to the $\ket{ 65\mathrm{s}, - }$ (continuous dark blue curve) and $\ket{ 66\mathrm{s}, + }$ (continuous light red curve) states asymptotically are shown in Fig.~\ref{fig:theory_figure}(a).
When compared to the diabatic potentials (dashed curves which neglect off-diagonal couplings) that cross at $R \approx 110$ nm,
the charge-dipole interaction induces a strong level repulsion in the adiabats between the $\ket{ 65\mathrm{s}, - }$ and $\ket{ 66\mathrm{s}, + }$ states [dark blue curve in Fig.~\ref{fig:theory_figure}(b)]. 
Collisional RET between the two adiabats occurs non-adiabatically over the interaction region (see Appendix~\ref{app:monopole_dipole}).

The transition rate $\Gamma(\ket{ \mathrm{i} } \rightarrow \ket{ \mathrm{f} })$ for RET between an initial (final) state $\ket{ \mathrm{i} } = \ket{ 65\mathrm{s}, - }$ ($\ket{ \mathrm{f} } = \ket{ 66\mathrm{s}, + }$), is obtained by considering the product of the NH$_3$ number density, $n_{\rm NH_3}$, and the thermally averaged rate constant 
$\beta( \ket{ \mathrm{i} } \rightarrow \ket{ \mathrm{f} } ) = \int ( \hbar K / \mu )\,\sigma_{ \ket{ \mathrm{i} } \rightarrow \ket{ \mathrm{f} } }\,c_{\rm MB}(\boldsymbol{K})\,\mathrm{d}^3\boldsymbol{K}$,
where $c_{\rm MB}(\boldsymbol{K})$ is a Maxwell-Boltzmann distribution of collision momenta. 
At collision energies much larger than the transition-state detuning $\Delta_{\mathrm{i},\mathrm{f}}$, the scattering cross section,
treated within the Born approximation \cite{Matsuzawa1979statechanging}, is
\begin{align} \label{eq:cross_section}
    \sigma_{ \ket{ \mathrm{i} } \rightarrow \ket{ \mathrm{f} } }
    &=
    \frac{ \pi }{ K_{\mathrm{i}}^2 }
    \abs{ 
    T_{ |{ \mathrm{i} }\rangle \rightarrow |{ \mathrm{f} }\rangle } }^2,
\end{align}
with the transition $T$-matrix 
\begin{align}
    T_{ |{ \mathrm{i} }\rangle \rightarrow |{ \mathrm{f} }\rangle }
    &\approx
    \frac{ 16 i \pi^2 \mu }{ \hbar^2 \sqrt{ K_{\mathrm{i}} K_{\mathrm{f}} } }
    \left(
    1 - \sqrt{ \frac{ 2 \mu \Delta_{\mathrm{i},\mathrm{f}} }{ \hbar^2 K_{\mathrm{i}}^2 } }
    \right) \\
    &\:\:\times 
    \int 
    \bra{ \mathrm{f} }
    V_{\rm CD}(R)
    \ket{ \mathrm{i} }
    \cos\left[
    \left( K_{\mathrm{f}} - K_{\mathrm{i}} \right) R
    \right]
    \,\mathrm{d}R, \nonumber
\end{align}
where $\mu$ is the reduced mass and $K_{\nu} = \sqrt{ 2 \mu ( E - \varepsilon_{\nu} ) / \hbar^2 }$ is the wavenumber corresponding to the collision threshold $\varepsilon_{\nu}$, with $\nu = \mathrm{i}, \mathrm{f}$ (see Supplemental Material for further details~\cite{SI}). 
The $T$-matrix element above implies that the efficacy of collisional RET is determined by the detuning of $K_{\mathrm{f}}$ from $K_{\mathrm{i}}$. Larger detunings result in more rapidly oscillating integrands, which suppress the RET cross section.

Considering the experimental conditions, for which the interaction time of the mixture of Rydberg He atoms and NH$_3$ molecules was $\Delta t = 12~\mu$s, the calculated probability $\mathbb{P}(|{ 65\mathrm{s},- }\rangle \rightarrow |{ 66\mathrm{s},+ }\rangle) = 1 - e^{-\Gamma \Delta t}$, that any given He atom undergoes RET to the $\ket{ 66\mathrm{s} }$ state is found to be $17\pm 4\%$. This is represented by the vertical bars in Fig.~\ref{fig2}(a).  
The uncertainty on this theoretical quantity results from propagation of the error in the number density and collision speed in the experiments.   
Similarly, we compute the non-resonant monopole-dipole induced population transfer between the $|{ 65\mathrm{s},+ }\rangle$ and $|{ 64\mathrm{s},- }\rangle$ states to be $\mathbb{P}\left( |{ 65\mathrm{s},+ }\rangle \rightarrow |{ 64\mathrm{s},- }\rangle \right) \approx 0.12\pm 0.03\%$,

Despite similar charge-dipole coupling strengths between the $|{ 65\mathrm{s},- }\rangle$ and $|{ 66\mathrm{s},+ }\rangle$ states, and  the $|{ 65\mathrm{s},+ }\rangle$ and $|{ 64\mathrm{s},- }\rangle$ states [Fig.~\ref{fig:theory_figure}(b)], the latter energy transfer process is highly suppressed because the $|{ 65\mathrm{s},+}\rangle\leftrightarrow|{ 64\mathrm{s},- }\rangle$ detuning of $\Delta_{\mathrm{i},\mathrm{f}}=1.166$~GHz is more than an order of magnitude larger than the $|{ 65\mathrm{s},- }\rangle\leftrightarrow|{ 66\mathrm{s},+ }\rangle$ detuning ($\Delta_{\mathrm{i},\mathrm{f}}=0.040$~GHz). Direct comparison of these results with the experimental data is seen in Fig.~\ref{fig2}(b). These data represent the difference between the normalized electric field ionization signals recorded following  preparation of the $|65\mathrm{s}\rangle$ state with and without NH$_3$. The dip observed close to 21~V/cm corresponds to the depletion of the $|65\mathrm{s}\rangle$ population by RET. The enhancement in the signal in the region labeled A, represents the resonant monopole-dipole energy transfer to the $|66\mathrm{s}\rangle$ state with the observed population transfer in excellent quantitative agreement with the calculations (thick vertical bar). The calculated value of $\mathbb{P}\left( |{ 65\mathrm{s},+ }\rangle \rightarrow |{ 64\mathrm{s},- }\rangle \right)$ is indicated by the small black bar labeled B at 22.5~V/cm. The calculated transfer in this case is below the experimental sensitivity. The theoretical treatment therefore corroborates the experimental observation of resonant monopole-dipole energy transfer. The resonant nature of this process, which is only seen at $n=65$ and~66, shows that the ionization signals do not arise as a result of short-range He$^+$-ion--dipole chemistry \cite{zhelyazkova21multipole}. 

\textit{Parity conservation} -- It is notable that the $\ket{ 65\mathrm{s},-}$ and $\ket{ 66\mathrm{s},+}$ pair states both comprise even parity Rydberg levels, while the inversion sublevels in NH$_3$ have opposite parity. 
The conservation of the total parity symmetry of the collision complex must, therefore, be enforced through the relative mechanical angular momentum~\cite{deiglmayr2014observation}. 
The requisite coupling between opposite parity collisional partial waves $\ket{L, m_L}$, occurs through the anisotropy of the charge-dipole interaction, which is proportional to $\cos{\theta_R}$, with $\theta_R$ the angle between the molecular dipole and collision coordinate \cite{SI}.
At the collision energies in the experiments, a large number of collisional partial waves ($L \approx 200$) are typically involved \cite{SI}, making it difficult to resolve the partial wave parity change. 
However, this may be observed in the future in differential scattering experiments in the ultracold regime close to the s-wave limit~\cite{aikawa14anisotropic, wang21anisotropic, li2021tuning}.

\textit{Conclusion} -- In summary, we have observed resonant monopole-dipole energy transfer between equal parity Rydberg levels in He, and the ground-state inversion doublet in NH$_3$. 
This resonant energy exchange is mediated by the charge-dipole interaction of the Rydberg electron with the polar molecule which leads to the hybridization of the opposite parity inversion sublevels in NH$_3$. Total parity is conserved by the admixture of collisional angular momenta of the atom-molecule complex.  
The theoretical treatment of the He($|n\mathrm{s}\rangle$)-NH$_3$ encounters, as charge-dipole--mediated scattering events, is in excellent quantitative agreement with the measured state-to-state population transfer.
The theory also provides insight into the suppression of non-resonant transition probabilities in such collisions.

Whereas all FRET processes identified up to now could be treated at far-field distances, we expect the resonant monopole-dipole energy transfer observed here to be a universal feature of charge-dipole interacting quantum systems in the near-field.
It is not restricted to He Rydberg atoms and NH$_3$ molecules, nor does it necessarily rely on the coupling to the molecule nuclear motion degree of freedom exploited here.
We envision that it could be observed in ultracold gases, expanding the range of tunable interactions available in hybrid atom-molecule platforms. Zeeman shifts of the Rydberg levels may be exploited to allow selected monopole transitions to be fined-tuned into resonance with an electric-dipole transition of the interacting partner without inducing longer-range interactions. Such control over the resonance condition presents resonant monopole-dipole energy transfer as a powerful tool for spin-motion coupled quantum applications \cite{mizrahi13ultrafast, dareau18observation, bharti24strong, picard2025entanglement, wang2025theory}.  
The association of Rydberg atoms and polar molecules into giant polyatomic Rydberg molecules \cite{rittenhouse2010ultracold, gonzalezferez2015rotational, bendkowsky2009observation} might also allow for continuous coherent oscillations between resonant Rydberg and molecular states, without the need for external confinement.
Moreover, recent advances in optical tweezer techniques to control and trap neutral atoms and molecules~\cite{anderegg2019optical, anderegg21observation, bluvstein2022quantum, ruttley2023formation, ruttley2024enhanced, ruttley2025long, guttridge2025individual} may permit on-demand access to monopole-dipole blockade~\cite{jaksch2000fast, lukin2001dipole} or antiblockade \cite{Ates07_PRL, Amthor10_PRL} physics for quantum simulation and computing.

\textit{Acknowledgements} -- This work is supported by the UK Science and Technology Facilities Research Council under Grant No. ST/T006439/1. RRWW and HRS acknowledge support at ITAMP from the National Science Foundation, and
 RGF acknowledges support from the Spanish project PID2023-147039NB-I00 (MICIN).

\bibliography{paperNotes}

@Article{cario1922uber,
author={Cario, G.
and Franck, J.},
title={{{\"U}ber Zerlegung von Wasserstoffmolek{\"u}len durch angeregte Quecksilberatome}},
journal={Z. Phys.},
year={1922},
month={Dec},
day={01},
volume={11},
number={1},
pages={161-166},
issn={0044-3328},
doi={10.1007/BF01328410},
url={https://doi.org/10.1007/BF01328410}
}

@article{perrin1927fluorescence,
  title={Fluorescence et induction mol{\'e}culaire par r{\'e}sonance},
  author={Perrin, Jean},
  journal={C. R. Hebd. S\'eances Acad. Sci.},
  volume={184},
  pages={1097--1100},
  year={1927}
}

@article{perrin32theory,
	author = {F. Perrin},
	title = {{Th\'eorie quantique des transferts d’activation entre mol\'ecules de m\^eme esp\`ece. Cas des solutions fluorescentes}},
	DOI= "10.1051/anphys/193210170283",
	url= "https://doi.org/10.1051/anphys/193210170283",
	journal = {Ann. Phys.},
	year = 1932,
	volume = 10,
	number = 17,
	pages = "283-314",
}

@article{kallmann1929on,
url = {https://doi.org/10.1515/zpch-1929-0214},
title = {{\"Uber quantenmechanische Energieübertragung zwischen atomaren Systemen}},
author = {H. Kallmann and F. London},
pages = {207--243},
volume = {2B},
number = {1},
journal = {Z. Phys. Chem.},
doi = {doi:10.1515/zpch-1929-0214},
year = {1929},
lastchecked = {2025-06-09}
}

@Article{forster1946energy,
author={Forster, Th.},
title={{Energiewanderung und Fluoreszenz}},
journal={Naturwissenschaften},
year={1946},
month={Jun},
day={01},
volume={33},
number={6},
pages={166-175},
issn={1432-1904},
doi={10.1007/BF00585226},
url={https://doi.org/10.1007/BF00585226}
}

@Article{mirkovic2017light,
author={Mirkovic, Tihana
and Ostroumov, Evgeny E.
and Anna, Jessica M.
and van Grondelle, Rienk
and {Govindjee}
and Scholes, Gregory D.},
title={Light Absorption and Energy Transfer in the Antenna Complexes of Photosynthetic Organisms},
journal={Chem. Rev.},
year={2017},
month={Jan},
day={25},
publisher={American Chemical Society},
volume={117},
number={2},
pages={249-293},
issn={0009-2665},
doi={10.1021/acs.chemrev.6b00002},
url={https://doi.org/10.1021/acs.chemrev.6b00002}
}

@ARTICLE{jones2019resonance,
author={Jones, Garth A.  and Bradshaw, David S. },
title={{Resonance Energy Transfer: From Fundamental Theory to Recent Applications}},
journal={Front. Phys.},
volume={7},
pages={100},
year={2019},
url={https://www.frontiersin.org/journals/physics/articles/10.3389/fphy.2019.00100},
doi={10.3389/fphy.2019.00100},
issn={2296-424X}
}

@article{sener2011forster,
author = {\c{S}ener, Melih and Str\"umpfer, Johan and Hsin, Jen and Chandler, Danielle and Scheuring, Simon and Hunter, C. Neil and Schulten, Klaus},
title = {F\"orster Energy Transfer Theory as Reflected in the Structures of Photosynthetic Light-Harvesting Systems},
journal = {ChemPhysChem},
volume = {12},
number = {3},
pages = {518-531},
year={2011},
doi = {https://doi.org/10.1002/cphc.201000944},
url = {https://chemistry-europe.onlinelibrary.wiley.com/doi/abs/10.1002/cphc.201000944},
}

@Article{ravets2014coherent,
author={Ravets, Sylvain
and Labuhn, Henning
and Barredo, Daniel
and B{\'e}guin, Lucas
and Lahaye, Thierry
and Browaeys, Antoine},
title={{Coherent dipole--dipole coupling between two single Rydberg atoms at an electrically-tuned F{\"o}rster resonance}},
journal={Nature Phys.},
year={2014},
month={Dec},
day={01},
volume={10},
number={12},
pages={914-917},
issn={1745-2481},
doi={10.1038/nphys3119},
url={https://doi.org/10.1038/nphys3119}
}

@article{deLeseleuc2017optical,
  title = {{Optical Control of the Resonant Dipole-Dipole Interaction between Rydberg Atoms}},
  author = {de~L\'es\'eleuc, Sylvain and Barredo, Daniel and Lienhard, Vincent and Browaeys, Antoine and Lahaye, Thierry},
  journal = {Phys. Rev. Lett.},
  volume = {119},
  issue = {5},
  pages = {053202},
  numpages = {6},
  year = {2017},
  month = {Aug},
  publisher = {American Physical Society},
  doi = {10.1103/PhysRevLett.119.053202},
  url = {https://link.aps.org/doi/10.1103/PhysRevLett.119.053202}
}

@article{aikawa14anisotropic,
  title = {{Anisotropic Relaxation Dynamics in a Dipolar Fermi Gas Driven Out of Equilibrium}},
  author = {Aikawa, K. and Frisch, A. and Mark, M. and Baier, S. and Grimm, R. and Bohn, J. L. and Jin, D. S. and Bruun, G. M. and Ferlaino, F.},
  journal = {Phys. Rev. Lett.},
  volume = {113},
  issue = {26},
  pages = {263201},
  numpages = {5},
  year = {2014},
  month = {Dec},
  publisher = {American Physical Society},
  doi = {10.1103/PhysRevLett.113.263201},
  url = {https://link.aps.org/doi/10.1103/PhysRevLett.113.263201}
}

@article{wang21anisotropic,
  title = {Anisotropic thermalization of dilute dipolar gases},
  author = {Wang, Reuben R. W. and Bohn, John L.},
  journal = {Phys. Rev. A},
  volume = {103},
  issue = {6},
  pages = {063320},
  numpages = {10},
  year = {2021},
  month = {Jun},
  publisher = {American Physical Society},
  doi = {10.1103/PhysRevA.103.063320},
  url = {https://link.aps.org/doi/10.1103/PhysRevA.103.063320}
}

@article{zhu2025probing,
  title = {{Probing Dipolar Interactions between Rydberg Atoms and Ultracold Polar Molecules}},
  author = {Zhu, Lingbang and Luke, Jeshurun and Shaham, Roy and Liu, Yi-Xiang and Ni, Kang-Kuen},
  journal = {Phys. Rev. Lett.},
  volume = {135},
  issue = {15},
  pages = {153001},
  numpages = {6},
  year = {2025},
  month = {Oct},
  publisher = {American Physical Society},
  doi = {10.1103/48rk-sxfs},
  url = {https://link.aps.org/doi/10.1103/48rk-sxfs}
}

@article{zhelyazkova2017electrically,
  title={Electrically tuned {F}{\"o}rster resonances in collisions of {NH$_3$} with {R}ydberg {H}e atoms},
  author={Zhelyazkova, V and Hogan, S D},
  journal={Phys. Rev. A},
  volume={95},
  number={4},
  pages={042710},
  year={2017},
  publisher={APS}
}

@article{smith78a,
  title = {{Discrete Energy Transfer in Collisions of $\mathrm{Xe}(n\mathrm{f})$ Rydberg Atoms with N${\mathrm{H}}_{3}$ Molecules}},
  author = {Smith, K. A. and Kellert, F. G. and Rundel, R. D. and Dunning, F. B. and Stebbings, R. F.},
  journal = {Phys. Rev. Lett.},
  volume = {40},
  issue = {21},
  pages = {1362--1365},
  numpages = {0},
  year = {1978},
  month = {May},
  publisher = {American Physical Society},
  doi = {10.1103/PhysRevLett.40.1362},
  url = {https://link.aps.org/doi/10.1103/PhysRevLett.40.1362}
}

@article{gawlas2019rydberg,
  title={Rydberg-state-resolved resonant energy transfer in cold electric-field-controlled intrabeam collisions of {NH$_3$} with {R}ydberg {H}e atoms},
  author={Gawlas, K and Hogan, S D},
  journal={J. Phys. Chem. Lett.},
  volume={11},
  number={1},
  pages={83--87},
  year={2019},
  publisher={ACS Publications}
}

@article{jarisch2018state,
doi = {10.1088/1367-2630/aaf02e},
url = {https://dx.doi.org/10.1088/1367-2630/aaf02e},
year = {2018},
month = {nov},
publisher = {IOP Publishing},
volume = {20},
number = {11},
pages = {113044},
author = {Jarisch, F and Zeppenfeld, M},
title = {{State resolved investigation of F\"orster resonant energy transfer in collisions between polar molecules and Rydberg atoms}},
journal = {New J. Phys.}
}

@Article{patsch2022rydberg,
author={Patsch, Sabrina
and Zeppenfeld, Martin
and Koch, Christiane P.},
title={{Rydberg Atom-Enabled Spectroscopy of Polar Molecules via F{\"o}rster Resonance Energy Transfer}},
journal={J. Phys. Chem. Lett.},
year={2022},
month={Nov},
day={24},
publisher={American Chemical Society},
volume={13},
number={46},
pages={10728-10733},
doi={10.1021/acs.jpclett.2c02521},
url={https://doi.org/10.1021/acs.jpclett.2c02521}
}

@article{zou2022probing,
  title={Probing van der {W}aals interactions and detecting polar molecules by {F}{\"o}rster-resonance energy transfer with {R}ydberg atoms at temperatures below 100 m{K}},
  author={Zou, J and Hogan, S D},
  journal={Phys. Rev. A},
  volume={106},
  number={4},
  pages={043111},
  year={2022},
  publisher={APS}
}

@article{Green80energy,
    author = {Green, Sheldon},
    title = {{Energy transfer in NH$_3$-He collisions}},
    journal = {J. Chem. Phys.},
    volume = {73},
    number = {6},
    pages = {2740-2750},
    year = {1980},
    month = {09},
    issn = {0021-9606},
    doi = {10.1063/1.440495},
    url = {https://doi.org/10.1063/1.440495},
}

@article{gallagher80resonant,
  title = {{Resonant electronic to vibrational energy transfer from Na to C${\mathrm{H}}_{4}$ and C${\mathrm{D}}_{4}$}},
  author = {Gallagher, T. F. and Ruff, G. A. and Safinya, K. A.},
  journal = {Phys. Rev. A},
  volume = {22},
  issue = {3},
  pages = {843--848},
  numpages = {0},
  year = {1980},
  month = {Sep},
  publisher = {American Physical Society},
  doi = {10.1103/PhysRevA.22.843},
  url = {https://link.aps.org/doi/10.1103/PhysRevA.22.843}
}

@article{deiglmayr2014observation,
  title = {{Observation of Dipole-Quadrupole Interaction in an Ultracold Gas of Rydberg Atoms}},
  author = {Deiglmayr, Johannes and Sa\ss{}mannshausen, Heiner and Pillet, Pierre and Merkt, Fr\'ed\'eric},
  journal = {Phys. Rev. Lett.},
  volume = {113},
  issue = {19},
  pages = {193001},
  numpages = {5},
  year = {2014},
  month = {Nov},
  publisher = {American Physical Society},
  doi = {10.1103/PhysRevLett.113.193001},
  url = {https://link.aps.org/doi/10.1103/PhysRevLett.113.193001}
}

@article{maineult2016dipole,
doi = {10.1088/0953-4075/49/21/214001},
url = {https://dx.doi.org/10.1088/0953-4075/49/21/214001},
year = {2016},
month = {oct},
publisher = {IOP Publishing},
volume = {49},
number = {21},
pages = {214001},
author = {Maineult, Wilfried and Pelle, Bruno and Faoro, Riccardo and Arimondo, Ennio and Pillet, Pierre and Cheinet, Patrick},
title = {{Dipole–quadrupole F\"orster resonance in cesium Rydberg gas}},
journal = {J. Phys. B: At. Mol. Opt. Phys.}
}

@article{carroll25observation,
author = {Annette N. Carroll  and Henrik Hirzler  and Calder Miller  and David Wellnitz  and Sean R. Muleady  and Junyu Lin  and Krzysztof P. Zamarski  and Reuben R. W. Wang  and John L. Bohn  and Ana Maria Rey  and Jun Ye },
title = {{Observation of generalized $t$-$J$ spin dynamics with tunable dipolar interactions}},
journal = {Science},
volume = {388},
number = {6745},
pages = {381-386},
year = {2025},
doi = {10.1126/science.adq0911},
URL = {https://www.science.org/doi/abs/10.1126/science.adq0911}
}

@article{wang22enriching,
  title = {{Enriching the Quantum Toolbox of Ultracold Molecules with Rydberg Atoms}},
  author = {Wang, Kenneth and Williams, Conner P. and Picard, Lewis R. B. and Yao, Norman Y. and Ni, Kang-Kuen},
  journal = {PRX Quantum},
  volume = {3},
  issue = {3},
  pages = {030339},
  numpages = {12},
  year = {2022},
  month = {Sep},
  publisher = {American Physical Society},
  doi = {10.1103/PRXQuantum.3.030339},
  url = {https://link.aps.org/doi/10.1103/PRXQuantum.3.030339}
}

@article{zhang22quantum,
  title = {{Quantum Computation in a Hybrid Array of Molecules and Rydberg Atoms}},
  author = {Zhang, Chi and Tarbutt, M. R.},
  journal = {PRX Quantum},
  volume = {3},
  issue = {3},
  pages = {030340},
  numpages = {17},
  year = {2022},
  month = {Sep},
  publisher = {American Physical Society},
  doi = {10.1103/PRXQuantum.3.030340},
  url = {https://link.aps.org/doi/10.1103/PRXQuantum.3.030340}
}

@article{simmons48structure,
  title = {{Structure of the Inversion Spectrum of Ammonia}},
  author = {Simmons, James W. and Gordy, Walter},
  journal = {Phys. Rev.},
  volume = {73},
  issue = {7},
  pages = {713--718},
  numpages = {0},
  year = {1948},
  month = {Apr},
  publisher = {American Physical Society},
  doi = {10.1103/PhysRev.73.713},
  url = {https://link.aps.org/doi/10.1103/PhysRev.73.713}
}

@article{halfmann2000source,
  title={A source for a high-intensity pulsed beam of metastable helium atoms},
  author={Halfmann, T and Koensgen, J and Bergmann, K},
  journal={Meas. Sci. Technol.},
  volume={11},
  number={10},
  pages={1510},
  year={2000},
  publisher={IOP Publishing}
}

@article{hogan2018laser,
  title={Laser photoexcitation of {R}ydberg states in helium with $n> 400$},
  author={Hogan, S D and Houston, Y and Wei, B},
  journal={J. Phys. B: At. Mol. Opt. Phys.},
  volume={51},
  number={14},
  pages={145002},
  year={2018},
  publisher={IOP Publishing}
}

@Article{zhelyazkova21multipole,
author = {Zhelyazkova, Valentina and Martins, Fernanda B. V. and Agner, Josef A. and Schmutz, Hansj\"urg and Merkt, Fr\'ed\'eric},
title  = {{Multipole-moment effects in ion–molecule reactions at low temperatures: part I – ion-dipole enhancement of the rate coefficients of the He$^+ + $NH$_3$ and He$^+ + $ND$_3$ reactions at collisional energies $E_{\mathrm{coll}}/k_{\mathrm{B}}$ near 0~K}},
journal  ="Phys. Chem. Chem. Phys.",
year  ="2021",
volume  ="23",
issue  ="38",
pages  ="21606-21622",
publisher  ="The Royal Society of Chemistry",
doi  ="10.1039/D1CP03116C",
url  ="http://dx.doi.org/10.1039/D1CP03116C",
}

@article{tsikritea22capture,
    author = {Tsikritea, Andriana and Diprose, Jake A. and Softley, Timothy P. and Heazlewood, Brianna R.},
    title = {Capture theory models: An overview of their development, experimental verification, and applications to ion–molecule reactions},
    journal = {J. Chem. Phys.},
    volume = {157},
    number = {6},
    pages = {060901},
    year = {2022},
    month = {08},
    issn = {0021-9606},
    doi = {10.1063/5.0098552},
    url = {https://doi.org/10.1063/5.0098552}
}

@article{Kuznetsova2016rydberg,
  title = {Rydberg-atom-mediated nondestructive readout of collective rotational states in polar-molecule arrays},
  author = {Kuznetsova, Elena and Rittenhouse, Seth T. and Sadeghpour, H. R. and Yelin, Susanne F.},
  journal = {Phys. Rev. A},
  volume = {94},
  issue = {3},
  pages = {032325},
  numpages = {21},
  year = {2016},
  month = {Sep},
  publisher = {American Physical Society},
  doi = {10.1103/PhysRevA.94.032325},
  url = {https://link.aps.org/doi/10.1103/PhysRevA.94.032325}
}

@article{Hickman1978theory,
  title = {{Theory of angular momentum mixing in Rydberg-atom-rare-gas collisions}},
  author = {Hickman, A. P.},
  journal = {Phys. Rev. A},
  volume = {18},
  issue = {4},
  pages = {1339--1342},
  numpages = {0},
  year = {1978},
  month = {Oct},
  publisher = {American Physical Society},
  doi = {10.1103/PhysRevA.18.1339},
  url = {https://link.aps.org/doi/10.1103/PhysRevA.18.1339}
}

@article{Hickman1979relation,
  title = {{Relation between low-energy-electron scattering and $\ell$-changing collisions of Rydberg atoms}},
  author = {Hickman, A. P.},
  journal = {Phys. Rev. A},
  volume = {19},
  issue = {3},
  pages = {994--1003},
  numpages = {0},
  year = {1979},
  month = {Mar},
  publisher = {American Physical Society},
  doi = {10.1103/PhysRevA.19.994},
  url = {https://link.aps.org/doi/10.1103/PhysRevA.19.994}
}

@article{Matsuzawa1979statechanging,
  title = {{State-changing collision of a high-Rydberg atom with a polar molecule}},
  author = {Matsuzawa, Michio},
  journal = {Phys. Rev. A},
  volume = {20},
  issue = {3},
  pages = {860--866},
  numpages = {0},
  year = {1979},
  month = {Sep},
  publisher = {American Physical Society},
  doi = {10.1103/PhysRevA.20.860},
  url = {https://link.aps.org/doi/10.1103/PhysRevA.20.860}
}

@article{Schlagmuller16ultracold,
  title = {{Ultracold Chemical Reactions of a Single Rydberg Atom in a Dense Gas}},
  author = {Schlagm\"uller, Michael and Liebisch, Tara Cubel and Engel, Felix and Kleinbach, Kathrin S. and B\"ottcher, Fabian and Hermann, Udo and Westphal, Karl M. and Gaj, Anita and L\"ow, Robert and Hofferberth, Sebastian and Pfau, Tilman and P\'erez-R\'{\i}os, Jes\'us and Greene, Chris H.},
  journal = {Phys. Rev. X},
  volume = {6},
  issue = {3},
  pages = {031020},
  numpages = {14},
  year = {2016},
  month = {Aug},
  publisher = {American Physical Society},
  doi = {10.1103/PhysRevX.6.031020},
  url = {https://link.aps.org/doi/10.1103/PhysRevX.6.031020}
}

@Article{Geppert2021diffusive,
author={Geppert, Philipp
and Alth{\"o}n, Max
and Fichtner, Daniel
and Ott, Herwig},
title={{Diffusive-like redistribution in state-changing collisions between Rydberg atoms and ground state atoms}},
journal={Nature Communications},
year={2021},
month={Jun},
day={23},
volume={12},
number={1},
pages={3900},
issn={2041-1723},
doi={10.1038/s41467-021-24146-0},
url={https://doi.org/10.1038/s41467-021-24146-0}
}

@Article{li2021tuning,
author={Li, Jun-Ru
and Tobias, William G.
and Matsuda, Kyle
and Miller, Calder
and Valtolina, Giacomo
and De Marco, Luigi
and Wang, Reuben R. W.
and Lassabli{\`e}re, Lucas
and Qu{\'e}m{\'e}ner, Goulven
and Bohn, John L.
and Ye, Jun},
title={Tuning of dipolar interactions and evaporative cooling in a three-dimensional molecular quantum gas},
journal={Nature Phys.},
year={2021},
month={Oct},
day={01},
volume={17},
number={10},
pages={1144-1148},
issn={1745-2481},
doi={10.1038/s41567-021-01329-6},
url={https://doi.org/10.1038/s41567-021-01329-6}
}

@article{lassabliere2022model,
  title = {{Model for two-body collisions between ultracold dipolar molecules around a F\"orster resonance in an electric field}},
  author = {Lassabli\`ere, Lucas and Qu\'em\'ener, Goulven},
  journal = {Phys. Rev. A},
  volume = {106},
  issue = {3},
  pages = {033311},
  numpages = {11},
  year = {2022},
  month = {Sep},
  publisher = {American Physical Society},
  doi = {10.1103/PhysRevA.106.033311},
  url = {https://link.aps.org/doi/10.1103/PhysRevA.106.033311}
}

@Article{ha2024flourescence,
author={Ha, Taekjip
and Fei, Jingyi
and Schmid, Sonja
and Lee, Nam Ki
and Gonzalez, Ruben L.
and Paul, Sneha
and Yeou, Sanghun},
title={Fluorescence resonance energy transfer at the single-molecule level},
journal={Nat. Rev. Methods Primers},
year={2024},
month={Mar},
day={28},
volume={4},
number={1},
pages={21},
issn={2662-8449},
doi={10.1038/s43586-024-00298-3},
url={https://doi.org/10.1038/s43586-024-00298-3}
}

@Article{yan2013observation,
author={Yan, Bo
and Moses, Steven A.
and Gadway, Bryce
and Covey, Jacob P.
and Hazzard, Kaden R. A.
and Rey, Ana Maria
and Jin, Deborah S.
and Ye, Jun},
title={Observation of dipolar spin-exchange interactions with lattice-confined polar molecules},
journal={Nature},
year={2013},
month={Sep},
day={01},
volume={501},
number={7468},
pages={521-525},
issn={1476-4687},
doi={10.1038/nature12483},
url={https://doi.org/10.1038/nature12483}
}

@article{vanDitzhuijzen2008spatially,
  title = {{Spatially Resolved Observation of Dipole-Dipole Interaction between Rydberg Atoms}},
  author = {van Ditzhuijzen, C. S. E. and Koenderink, A. F. and Hern\'andez, J. V. and Robicheaux, F. and Noordam, L. D. and van~Linden van~den~Heuvell, H. B.},
  journal = {Phys. Rev. Lett.},
  volume = {100},
  issue = {24},
  pages = {243201},
  numpages = {4},
  year = {2008},
  month = {Jun},
  publisher = {American Physical Society},
  doi = {10.1103/PhysRevLett.100.243201},
  url = {https://link.aps.org/doi/10.1103/PhysRevLett.100.243201}
}

@Article{christakis2023probing,
author={Christakis, Lysander
and Rosenberg, Jason S.
and Raj, Ravin
and Chi, Sungjae
and Morningstar, Alan
and Huse, David A.
and Yan, Zoe Z.
and Bakr, Waseem S.},
title={Probing site-resolved correlations in a spin system of ultracold molecules},
journal={Nature},
year={2023},
month={Feb},
day={01},
volume={614},
number={7946},
pages={64-69},
issn={1476-4687},
doi={10.1038/s41586-022-05558-4},
url={https://doi.org/10.1038/s41586-022-05558-4}
}

@article{jaksch2000fast,
  title = {{Fast Quantum Gates for Neutral Atoms}},
  author = {Jaksch, D. and Cirac, J. I. and Zoller, P. and Rolston, S. L. and C\^ot\'e, R. and Lukin, M. D.},
  journal = {Phys. Rev. Lett.},
  volume = {85},
  issue = {10},
  pages = {2208--2211},
  numpages = {0},
  year = {2000},
  month = {Sep},
  publisher = {American Physical Society},
  doi = {10.1103/PhysRevLett.85.2208},
  url = {https://link.aps.org/doi/10.1103/PhysRevLett.85.2208}
}

@article{lukin2001dipole,
  title = {{Dipole Blockade and Quantum Information Processing in Mesoscopic Atomic Ensembles}},
  author = {Lukin, M. D. and Fleischhauer, M. and Cote, R. and Duan, L. M. and Jaksch, D. and Cirac, J. I. and Zoller, P.},
  journal = {Phys. Rev. Lett.},
  volume = {87},
  issue = {3},
  pages = {037901},
  numpages = {4},
  year = {2001},
  month = {Jun},
  publisher = {American Physical Society},
  doi = {10.1103/PhysRevLett.87.037901},
  url = {https://link.aps.org/doi/10.1103/PhysRevLett.87.037901}
}

@article{guttridge23observation,
  title = {{Observation of Rydberg Blockade Due to the Charge-Dipole Interaction between an Atom and a Polar Molecule}},
  author = {Guttridge, Alexander and Ruttley, Daniel K. and Baldock, Archie C. and Gonz\'alez-F\'erez, Rosario and Sadeghpour, H. R. and Adams, C. S. and Cornish, Simon L.},
  journal = {Phys. Rev. Lett.},
  volume = {131},
  issue = {1},
  pages = {013401},
  numpages = {8},
  year = {2023},
  month = {Jul},
  publisher = {American Physical Society},
  doi = {10.1103/PhysRevLett.131.013401},
  url = {https://link.aps.org/doi/10.1103/PhysRevLett.131.013401}
}

@article{Ates07_PRL,
  title = {{Antiblockade in Rydberg Excitation of an Ultracold Lattice Gas}},
  author = {Ates, C. and Pohl, T. and Pattard, T. and Rost, J. M.},
  journal = {Phys. Rev. Lett.},
  volume = {98},
  issue = {2},
  pages = {023002},
  numpages = {4},
  year = {2007},
  month = {Jan},
  publisher = {American Physical Society},
  doi = {10.1103/PhysRevLett.98.023002},
  url = {https://link.aps.org/doi/10.1103/PhysRevLett.98.023002}
}

@article{Amthor10_PRL,
  title = {{Evidence of Antiblockade in an Ultracold Rydberg Gas}},
  author = {Amthor, Thomas and Giese, Christian and Hofmann, Christoph S. and Weidem\"uller, Matthias},
  journal = {Phys. Rev. Lett.},
  volume = {104},
  issue = {1},
  pages = {013001},
  numpages = {4},
  year = {2010},
  month = {Jan},
  publisher = {American Physical Society},
  doi = {10.1103/PhysRevLett.104.013001},
  url = {https://link.aps.org/doi/10.1103/PhysRevLett.104.013001}
}

@Article{picard2025entanglement,
author={Picard, Lewis R. B.
and Park, Annie J.
and Patenotte, Gabriel E.
and Gebretsadkan, Samuel
and Wellnitz, David
and Rey, Ana Maria
and Ni, Kang-Kuen},
title={{Entanglement and iSWAP gate between molecular qubits}},
journal={Nature},
year={2025},
month={Jan},
day={01},
volume={637},
number={8047},
pages={821-826},
issn={1476-4687},
doi={10.1038/s41586-024-08177-3},
url={https://doi.org/10.1038/s41586-024-08177-3}
}

@article{anderegg2019optical,
author = {Lo\"ic Anderegg  and Lawrence W. Cheuk  and Yicheng Bao  and Sean Burchesky  and Wolfgang Ketterle  and Kang-Kuen Ni  and John M. Doyle },
title = {An optical tweezer array of ultracold molecules},
journal = {Science},
volume = {365},
number = {6458},
pages = {1156-1158},
year = {2019},
doi = {10.1126/science.aax1265},
URL = {https://www.science.org/doi/abs/10.1126/science.aax1265},
}

@article{anderegg21observation,
author = {Lo\"ic Anderegg  and Sean Burchesky  and Yicheng Bao  and Scarlett S. Yu  and Tijs Karman  and Eunmi Chae  and Kang-Kuen Ni  and Wolfgang Ketterle  and John M. Doyle },
title = {Observation of microwave shielding of ultracold molecules},
journal = {Science},
volume = {373},
number = {6556},
pages = {779-782},
year = {2021},
doi = {10.1126/science.abg9502},
URL = {https://www.science.org/doi/abs/10.1126/science.abg9502}
}

@Article{bluvstein2022quantum,
author={Bluvstein, Dolev
and Levine, Harry
and Semeghini, Giulia
and Wang, Tout T.
and Ebadi, Sepehr
and Kalinowski, Marcin
and Keesling, Alexander
and Maskara, Nishad
and Pichler, Hannes
and Greiner, Markus
and Vuleti{\'{c}}, Vladan
and Lukin, Mikhail D.},
title={A quantum processor based on coherent transport of entangled atom arrays},
journal={Nature},
year={2022},
month={Apr},
day={01},
volume={604},
number={7906},
pages={451-456},
issn={1476-4687},
doi={10.1038/s41586-022-04592-6},
url={https://doi.org/10.1038/s41586-022-04592-6}
}

@article{ruttley2023formation,
  title = {{Formation of Ultracold Molecules by Merging Optical Tweezers}},
  author = {Ruttley, Daniel K. and Guttridge, Alexander and Spence, Stefan and Bird, Robert C. and Le Sueur, C. Ruth and Hutson, Jeremy M. and Cornish, Simon L.},
  journal = {Phys. Rev. Lett.},
  volume = {130},
  issue = {22},
  pages = {223401},
  numpages = {7},
  year = {2023},
  month = {May},
  publisher = {American Physical Society},
  doi = {10.1103/PhysRevLett.130.223401},
  url = {https://link.aps.org/doi/10.1103/PhysRevLett.130.223401}
}

@article{ruttley2024enhanced,
  title = {{Enhanced Quantum Control of Individual Ultracold Molecules Using Optical Tweezer Arrays}},
  author = {Ruttley, Daniel K. and Guttridge, Alexander and Hepworth, Tom R. and Cornish, Simon L.},
  journal = {PRX Quantum},
  volume = {5},
  issue = {2},
  pages = {020333},
  numpages = {19},
  year = {2024},
  month = {May},
  publisher = {American Physical Society},
  doi = {10.1103/PRXQuantum.5.020333},
  url = {https://link.aps.org/doi/10.1103/PRXQuantum.5.020333}
}

@Article{ruttley2025long,
author={Ruttley, Daniel K.
and Hepworth, Tom R.
and Guttridge, Alexander
and Cornish, Simon L.},
title={Long-lived entanglement of molecules in magic-wavelength optical tweezers},
journal={Nature},
year={2025},
month={Jan},
day={01},
volume={637},
number={8047},
pages={827-832},
issn={1476-4687},
doi={10.1038/s41586-024-08365-1},
url={https://doi.org/10.1038/s41586-024-08365-1}
}

@article{bharti24strong,
  title = {{Strong Spin-Motion Coupling in the Ultrafast Dynamics of Rydberg Atoms}},
  author = {Bharti, V. and Sugawa, S. and Kunimi, M. and Chauhan, V. S. and Mahesh, T. P. and Mizoguchi, M. and Matsubara, T. and Tomita, T. and de L\'es\'eleuc, S. and Ohmori, K.},
  journal = {Phys. Rev. Lett.},
  volume = {133},
  issue = {9},
  pages = {093405},
  numpages = {7},
  year = {2024},
  month = {Aug},
  publisher = {American Physical Society},
  doi = {10.1103/PhysRevLett.133.093405},
  url = {https://link.aps.org/doi/10.1103/PhysRevLett.133.093405}
}

@article{dareau18observation,
  title = {{Observation of Ultrastrong Spin-Motion Coupling for Cold Atoms in Optical Microtraps}},
  author = {Dareau, A. and Meng, Y. and Schneeweiss, P. and Rauschenbeutel, A.},
  journal = {Phys. Rev. Lett.},
  volume = {121},
  issue = {25},
  pages = {253603},
  numpages = {6},
  year = {2018},
  month = {Dec},
  publisher = {American Physical Society},
  doi = {10.1103/PhysRevLett.121.253603},
  url = {https://link.aps.org/doi/10.1103/PhysRevLett.121.253603}
}

@article{mizrahi13ultrafast,
  title = {{Ultrafast Spin-Motion Entanglement and Interferometry with a Single Atom}},
  author = {Mizrahi, J. and Senko, C. and Neyenhuis, B. and Johnson, K. G. and Campbell, W. C. and Conover, C. W. S. and Monroe, C.},
  journal = {Phys. Rev. Lett.},
  volume = {110},
  issue = {20},
  pages = {203001},
  numpages = {5},
  year = {2013},
  month = {May},
  publisher = {American Physical Society},
  doi = {10.1103/PhysRevLett.110.203001},
  url = {https://link.aps.org/doi/10.1103/PhysRevLett.110.203001}
}

@article{wang2025theory,
  title = {Theory of itinerant collisional spin dynamics in nondegenerate molecular gases},
  author = {Wang, Reuben R. W. and Bohn, John L.},
  journal = {Phys. Rev. A},
  volume = {112},
  issue = {4},
  pages = {043315},
  numpages = {20},
  year = {2025},
  month = {Oct},
  publisher = {American Physical Society},
  doi = {10.1103/bsfk-57cg},
  url = {https://link.aps.org/doi/10.1103/bsfk-57cg}
}

@article{rittenhouse2010ultracold,
  title = {{Ultracold Giant Polyatomic Rydberg Molecules: Coherent Control of Molecular Orientation}},
  author = {Rittenhouse, Seth T. and Sadeghpour, H. R.},
  journal = {Phys. Rev. Lett.},
  volume = {104},
  issue = {24},
  pages = {243002},
  numpages = {4},
  year = {2010},
  month = {Jun},
  publisher = {American Physical Society},
  doi = {10.1103/PhysRevLett.104.243002},
  url = {https://link.aps.org/doi/10.1103/PhysRevLett.104.243002}
}

@article{gonzalezferez2015rotational,
doi = {10.1088/1367-2630/17/1/013021},
url = {https://dx.doi.org/10.1088/1367-2630/17/1/013021},
year = {2015},
month = {jan},
publisher = {IOP Publishing},
volume = {17},
number = {1},
pages = {013021},
author = {González-Férez, Rosario and Sadeghpour, H R and Schmelcher, Peter},
title = {{Rotational hybridization, and control of alignment and orientation in triatomic ultralong-range Rydberg molecules}},
journal = {New J. Phys.},
}

@Article{bendkowsky2009observation,
author={Bendkowsky, Vera
and Butscher, Bj{\"o}rn
and Nipper, Johannes
and Shaffer, James P.
and L{\"o}w, Robert
and Pfau, Tilman},
title={{Observation of ultralong-range Rydberg molecules}},
journal={Nature},
year={2009},
month={Apr},
day={01},
volume={458},
number={7241},
pages={1005-1008},
issn={1476-4687},
doi={10.1038/nature07945},
url={https://doi.org/10.1038/nature07945}
}

@article{guttridge2025individual,
  title = {{Individual Assembly of Two-Species Rydberg Molecules Using Optical Tweezers}},
  author = {Guttridge, Alexander and Hepworth, Tom R. and Ruttley, Daniel K. and Durst, Aileen A. T. and Eiles, Matthew T. and Cornish, Simon L.},
  journal = {Phys. Rev. Lett.},
  volume = {134},
  issue = {13},
  pages = {133401},
  numpages = {7},
  year = {2025},
  month = {Apr},
  publisher = {American Physical Society},
  doi = {10.1103/PhysRevLett.134.133401},
  url = {https://link.aps.org/doi/10.1103/PhysRevLett.134.133401}
}

@misc{SI,
  note = {See Supplemental Material, which includes Refs. [63-68]}
}

@article{fermi47,
  title = {{The Capture of Negative Mesotrons in Matter}},
  author = {Fermi, E. and Teller, E.},
  journal = {Phys. Rev.},
  volume = {72},
  issue = {5},
  pages = {399--408},
  numpages = {0},
  year = {1947},
  month = {Sep},
  publisher = {American Physical Society},
  doi = {10.1103/PhysRev.72.399},
  url = {https://link.aps.org/doi/10.1103/PhysRev.72.399}
}

@article{Drake_1999,
doi = {10.1238/Physica.Topical.083a00083},
url = {https://dx.doi.org/10.1238/Physica.Topical.083a00083},
year = {1999},
month = {jan},
publisher = {},
volume = {1999},
number = {T83},
pages = {83},
author = {G W F Drake},
title = {High Precision Theory of Atomic Helium},
journal = {Phys. Scr.},
}

@article{church56a,
  title = {Electric-Monopole Transitions in Atomic Nuclei},
  author = {Church, E. L. and Weneser, J.},
  journal = {Phys. Rev.},
  volume = {103},
  issue = {4},
  pages = {1035--1044},
  numpages = {0},
  year = {1956},
  month = {Aug},
  publisher = {American Physical Society},
  doi = {10.1103/PhysRev.103.1035},
}

@article{zerguine08a,
  title = {Correlating Radii and Electric Monopole Transitions of Atomic Nuclei},
  author = {Zerguine, S. and Van Isacker, P. and Bouldjedri, A. and Heinze, S.},
  journal = {Phys. Rev. Lett.},
  volume = {101},
  issue = {2},
  pages = {022502},
  numpages = {4},
  year = {2008},
  month = {Jul},
  publisher = {American Physical Society},
  doi = {10.1103/PhysRevLett.101.022502},
}

\setcounter{secnumdepth}{2}
\begin{appendix}

\section{Charge-dipole mediated resonant monopole-dipole transition \label{app:monopole_dipole} }

The resonant monopole-dipole energy transfer reported here has, to our knowledge, not been previously described, warranting further discussion. 
With a more comprehensive model provided in the Supplementary Material \cite{SI}, we present an effective two-channel model in this appendix to elucidate the core physics of the RET process.
During a two-body collision, the Hamiltonian in the center of mass frame of the collision-complex, is
\begin{align} 
    H
    &=
    H_0 + V_{\rm CD}(\boldsymbol{r}, \boldsymbol{R}) \nonumber\\
    &=
    H_{\rm He}
    +
    H_{\rm NH_3}
    +
    \frac{ \boldsymbol{P}^2 }{ 2 \mu }
    +
    V_{\rm CD}(\boldsymbol{r}, \boldsymbol{R}),
\end{align}
where $\boldsymbol{P}$ is the relative momentum of the He atom and the NH$_3$ molecule, and $\mu = m_{\rm He} m_{\rm NH_3} / ( m_{\rm He} + m_{\rm NH_3} )$ is the reduced mass of the collision partners. 
In the combined atom-molecule Hilbert space, it is most natural to employ the basis in which the non-interacting Hamiltonian $(H_{\rm He} + H_{\rm NH_3}) \ket{n \ell m_{\ell}}\ket{\Pi} = E_{\Pi}^{n \ell m_{\ell}} \ket{n \ell m_{\ell}}\ket{\Pi}$, where $E_{\Pi}^{n \ell m_{\ell}}$, the sum of corresponding non-interacting He and NH$_3$ energies is diagonal, to construct a matrix representation of $V_{\rm CD}(\boldsymbol{R})$. 
In doing this, we ignore the molecular rotational structure by virtue of an adiabatic approximation (see \cite{SI} and references therein for further details).

In this situation, because all other states are energetically much further detuned compared to their splitting, we adopt a minimal model by truncating our basis set to just the following two states: 
\begin{subequations}
\begin{align}
    \ket{ \downarrow }
    &=
    \ket{65\mathrm{s}, -}, \\
    \ket{ \uparrow }
    &=
    \ket{66\mathrm{s}, +}.
\end{align}
\end{subequations}
The resulting interaction matrix is given by
\begin{align}
    \boldsymbol{V}_{\rm CD}(\boldsymbol{R})
    &=
    \begin{pmatrix}
        \bra{ \downarrow } {V}_{\rm CD}(\boldsymbol{R}) \ket{ \downarrow } &
        \bra{ \downarrow } {V}_{\rm CD}(\boldsymbol{R}) \ket{ \uparrow } \\
        \bra{ \uparrow } {V}_{\rm CD}(\boldsymbol{R}) \ket{ \downarrow } & 
        \bra{ \uparrow } {V}_{\rm CD}(\boldsymbol{R}) \ket{ \uparrow }
    \end{pmatrix},
\end{align}
with matrix elements
\begin{align}
    & \bra{ \Pi' }\bra{ n' \mathrm{s} }
    V_{\rm CD}(\boldsymbol{R})
    \ket{ n \mathrm{s}}\ket{ \Pi } \\
    &\approx
    -\frac{ e \bra{\Pi'} \mu_{\rm NH_3} \ket{\Pi} }{ R^{2} }
    \int_{0}^{R} r^{2} \mathrm{d}r
    {\cal R}_{n' \mathrm{s}}(r){\cal R}_{n \mathrm{s}}(r)
    \cos\theta_R, \nonumber
\end{align}
as shown in Fig.~\ref{fig:theory_figure}, 
where $\theta_R = \cos^{-1}( \hat{\boldsymbol{\mu}}_{\rm NH_3} \cdot \hat{\boldsymbol{R}} )$ is the angle between the molecular dipole and collision axis, and ${\cal R}_{n \mathrm{s}}(r)$ are the Rydberg state radial wavefunctions. 
Importantly, while the radial wavefunctions are orthogonal over the entire radial domain $\int_0^{\infty} {\cal R}_{n' \ell'}(r) {\cal R}_{n \ell}(r) r^2 \mathrm{d}r = \delta_{n', n} \delta_{\ell', \ell}$, the overlap for $|n'\mathrm{s}\rangle$ and $|n\mathrm{s}\rangle$ states is generally nonzero over a finite region of $r < R$.
This point makes clear the necessity for the molecule to enter the Rydberg electron charge distribution for resonant monopole-dipole energy transfer to occur. Under these conditions, for the triplet $\ell=0$ Rydberg states in He with an energy splitting which is close to resonance with the NH$_3$ inversion interval, the monopole term in the expansion, for which $r<R$, dominates. When integrated over this finite range, up to the position of the molecule, this term is non-zero.

Finally, regarding the scattering process, we can further expand the collisional angular momentum in partial waves $\ket{ L, m_L }$:
\begin{align}
    & \bra{ L', m'_L }\bra{ \Pi' }\bra{ n' \mathrm{s} }
    V_{\rm CD}(\boldsymbol{R})
    \ket{ n \mathrm{s}}\ket{ \Pi }
    \ket{ L, m_L } \nonumber\\
    &\approx
    -\frac{ e \bra{\Pi'} \mu_{\rm NH_3} \ket{\Pi} }{ R^{2} }
    \int_{0}^{R} r^{2} \mathrm{d}r
    {\cal R}_{n' \mathrm{s}}(r){\cal R}_{n \mathrm{s}}(r) \nonumber\\
    &\quad\quad \times 
    (-1)^{m'_L}
    \sqrt{ (2 L' + 1) (2 L + 1) } \nonumber\\
    &\quad\quad \times 
    \begin{pmatrix}
        L' & 1 & L \\
        -m'_L & 0 & m_L 
    \end{pmatrix}
    \begin{pmatrix}
        L' & 1 & L \\
        0 & 0 & 0 
    \end{pmatrix},
\end{align}
where the $2 \times 3$ arrays denote Wigner 3-j symbols. This expansion makes it clear, through the selection rule that $L' = L \pm 1$, that a collision must induce a single-quanta change of collisional angular momentum for each partial wave, changing its parity along with the parity of the NH$_3$ inversion sublevel.   
The total parity of the collision complex therefore remains conserved during the resonant monopole-dipole energy transfer process.

\section{Experimental details \label{app:expt} }

The conditions under which the experiments described here were performed were similar to those employed previously to study resonant dipole-dipole interactions between Rydberg He atoms and NH$_3$ in Ref.~\cite{gawlas2019rydberg,zou2022probing}. In the present experiments, the region in which the atoms were laser photoexcited to Rydberg states, interacted with the molecules, and were detected by state-selective electric field ionization was located between two parallel copper electrodes separated by 1.43~cm.

To measure and minimize stray electric fields in the region between these electrodes, microwave spectra of the single-color two-photon $|65\rm{s}\rangle\rightarrow|63\rm{s}\rangle$ transition were recorded with a pure beam of He. In this process, 2-$\mu$s-duration pulses of microwave radiation were applied 1, 3, or 5~$\mu$s after Rydberg state photoexcitation, before detection by electric field ionization. These spectra were recorded with a range of offset potentials applied to one of the copper electrodes. The offset potential required to minimize the Stark shift of the transition was determined from these measurements, and then applied to compensate the stray field. For the pure beam of He atoms, the residual Stark shift of the $|65\mathrm{s}\rangle\rightarrow|63\mathrm{s}\rangle$ transition under these conditions indicated that the remaining uncancelled stray electric field was $\sim13$~mV/cm.

In the measurements presented in Fig.~\ref{fig3} to state-selectively probe atoms in the $|66\mathrm{s}\rangle$ state that had undergone resonant energy transfer in a collision with an NH$_3$ molecule, the slightly larger residual stray electric field of $16.5\pm4.5$~mV/cm was determined with the molecules present in the beam. This stray field mixes $<0.03\%$ of $|n\mathrm{p}\rangle$ character into the $|n\mathrm{s}\rangle$ states in He with values of $n$ close to 65. Since, for example, the single-photon $|65\mathrm{s}\rangle\rightarrow|66\mathrm{p}\rangle$ and $|65\mathrm{p}\rangle\rightarrow|66\mathrm{s}\rangle$ transitions have electric dipole transition moments of $115\,e\,a_0$ (295~D) and $1655\,e\,a_0$ (4205~D), respectively, the dominant contribution to an electric dipole transition moment between the $\ell$-mixed $|65\mathrm{s}'\rangle$ and $|66\mathrm{s}'\rangle$ states in a weak stray electric field results from the $|65\mathrm{p}\rangle$ amplitude of $\sqrt{0.0003} = 0.017$ in the lower state. This yields an electric dipole moment for the single-photon $|65\mathrm{s}'\rangle\rightarrow|66\mathrm{s}'\rangle$ transition in a 16~mV/cm electric field of $\sim30\,e\,a_0$ (75~D).

From the work reported in Ref.~\cite{zou2022probing}, we determined that an electric dipole transition moment of $1285\,e\,a_0$ (3265~D) resulted in $\sim10$\% population transfer from the 40s state to the 40p state as a result of the resonant dipole-dipole interaction with the NH$_3$ molecules. In the work reported here on resonant monopole-dipole energy transfer at similar collision speeds and for comparable molecule number densities, we observe $\sim17\%$ population transfer. Because the electric dipole transition moments induced by $\ell$-mixing in the weak residual stray electric field in the experiments are $<2.5\%$ of those encountered in the studies of resonant dipole-dipole interactions at lower values of $n$, they cannot be primarily responsible for the energy transfer observed here.
    
\end{appendix}

\allowdisplaybreaks
\setcounter{section}{0}

\begin{center}
{\Large Supplemental Material}\\
\vspace*{0.5cm}
{\large for ‘Observation of resonant monopole-dipole energy transfer between Rydberg atoms and polar molecules'}\\
\vspace*{0.2cm}
J. Zou, R. R. W. Wang, R. Gonz\'alez-F\'erez, H.~R.~Sadeghpour and S. D. Hogan
\end{center}
\setcounter{page}{1}

\section{ Collision theory of charge-dipole mediated RET }

In this supplemental material, we study the collision dynamics of He atoms prepared in the $\ket{n, \ell, m_{\ell}} = \ket{65, 0, 0} \equiv \ket{ 65\mathrm{s} }$ Rydberg state, with NH$_3$ molecules in their $\ket{ - }$ inversion state of electronic ground state. 

\subsection{ Calculation of the diabatic matrix elements }

The Born-Oppenheimer Hamiltonian at large separations between the He atom and the NH$_3$ molecule is~\cite{rittenhouse2010ultracold, gonzalezferez2015rotational}
\begin{equation}
\label{eq:Hamil_adiabatic}
H=H_{\text{He}}+ H_{\text{NH}_3}+V_{\mathrm{CD}}(\mathbf{r},\mathbf{R}) 
\end{equation}
where $\mathbf{r}$ and $\mathbf{R}$ are the positions of the Rydberg electron and 
molecule with respect to the He$^+$ ion core, respectively.  
$H_{\text{He}}$ represents the single-electron Hamiltonian describing the Rydberg He atom,
and $H_{\text{NH}_3}=E_+|+\rangle\langle\, +\,|+E_-|-\rangle\langle -|$
the  Hamiltonian of NH$_3$, which is described as a two-state
parity doublet with $(E_{-} - E_{+})/h=23.695$~GHz.
The charge-dipole term, in Eq.~(\ref{eq:charge_dipole}) of the main text,  accounts for scattering of the electron from a molecule possessing a permanent electric dipole moment below a critical value $\mu_{\mathrm{cr}} = 1.639$~D~\cite{fermi47}; for supercritical dipole moments, the Rydberg electron could become bound to the molecule.

The Schr\"odinger equation associated with the Hamiltonian in Eq.~\eqref{eq:Hamil_adiabatic}
is solved by expanding the wave function in a basis set, as 
\begin{equation}
\label{eq:basis_exp}
\ket{\Psi}=\sum_{n,\ell,m_{\ell},\Pi} 
 C_{n\ell m_{\ell}\Pi}(\mathbf{R}) \ket{n\ell m_{\ell}}\ket{\Pi}, 
\end{equation}
where $\ket{\Pi}$, $\Pi=+,-$, are the inversion sublevels of NH$_3$, and 
$\ket{n\ell m_{\ell}}$ is the Rydberg electron wave function
with $n$, $\ell$, and $m_{\ell}$ being the principal, orbital angular momentum, and magnetic quantum numbers, respectively.
For He, the wave functions are computed using the Rydberg quantum defects
of Ref.~\cite{Drake_1999}. 
For our calculations, we include the Rydberg states of He with $n=64,65$ and $66$, and orbital angular
momentum $\ell\le 14$.
Due to the resonant condition of the $\ket{ 65\mathrm{s},- }$ and $\ket{ 66\mathrm{s},+ }$ 
states in He-NH$_3$, they have the largest weights,
$|C_{65\mathrm{s},-}(\mathbf{R})|^2$ and $|C_{66\mathrm{s},+}(\mathbf{R})|^2$,
in the total wave function~\eqref{eq:basis_exp}. The He Rydberg states with $\ell\geq 1$ do not contribute significantly. 
We find that the sum of weights over all other energetically nearby states satisfies $\sum_{n=64,65,66} \sum_{l\geq 1} \sum_{\Pi=+,-} |C_{n\ell m_{\ell}\Pi}(\boldsymbol{R})|^2 < 0.002$ for $R \leq 100$~nm.
\\

\subsection{ Resonant charge-dipole matrix element }

For completeness, we include the charge-dipole expansion used in our calculations, along with the coupling matrix element relevant for the resonant interactions of interest. 
For convenience, and without loss of generality, we consider the interactions in the collision-frame defined by $\hat{\boldsymbol{R}} = \hat{\boldsymbol{z}}$.
In this frame, the interaction of the molecule with the ion core is just $V^{{\rm He}^+}_{\rm CD} = {e d_{z} / R^{2}}$, while the electron-molecule terms are given by \cite{zhelyazkova2017electrically}
\begin{widetext}
\begin{align}
    V_{\rm CD}^{e^-}
    &=
    -\frac{ e \mu_{\rm NH_3}^- }{ 2 R^2 }
    \begin{cases}
        \sum_{\ell''=1}^{\infty}\sqrt{\frac{4\pi \ell''(\ell''+1)}{2\ell''+1}}\frac{r^{\ell''}}{R^{\ell''}}Y_{\ell''}^{1}(\theta,\phi), \quad& r< R, \\
        \sum_{\ell''=1}^{\infty}\sqrt{\frac{4\pi \ell''(\ell''+1)}{2\ell''+1}}\frac{R^{\ell''+1}}{r^{\ell''+1}}Y_{\ell''}^{1}(\theta,\phi), \quad& r > R 
    \end{cases} \nonumber\\
    &\quad 
    +
    \frac{e \mu_{\rm NH_3}^+ }{ 2 R^2 }
    \begin{cases}
        \sum_{\ell''=1}^{\infty}\sqrt{\frac{4\pi \ell''(\ell''+1)}{2\ell''+1}}\frac{r^{\ell''}}{R^{\ell''}}Y_{\ell''}^{-1}(\theta,\phi), \quad& r < R, \\
        \sum_{\ell''=1}^{\infty}\sqrt{\frac{4\pi \ell''(\ell''+1)}{2\ell''+1}}\frac{R^{\ell''+1}}{r^{\ell''+1}}Y_{\ell''}^{-1}(\theta,\phi), \quad& r > R 
    \end{cases} \nonumber\\
    &\quad 
    +
    \frac{ e \mu_{\rm NH_3}^z }{ R^2 } 
    \begin{cases}
        -\sum_{\ell''=0}^{\infty} \sqrt{\frac{4\pi}{2\ell''+1}}(\ell''+1)\frac{r^{\ell''}}{R^{\ell''}}Y_{\ell''}^{0}(\theta,\phi), \quad& r < R, \nonumber \\
        \sum_{\ell''=0}^{\infty}\sqrt{\frac{4\pi}{2\ell''+1}}\ell''\frac{R^{\ell''+1}}{r^{\ell''+1}}Y_{\ell''}^{0}(\theta,\phi), \quad& r > R, 
    \end{cases}
\end{align}
\end{widetext}
where $\mu_{\rm NH_3}^i$ are the $i$-th components of the molecular dipole vector with $\mu_{\rm NH_3}^{\pm} = \mu_{\rm NH_3}^{x} \pm i \mu_{\rm NH_3}^{y}$.  
For the resonant states of interest, we find that only the monopole ($\ell''=0$) term in the expansion is relevant, isolating the dominant coupling matrix element:
\begin{align}
    & \bra{ n' \mathrm{s}, + }
    V_{\rm CD}(R)
    \ket{ n \mathrm{s}, - } \nonumber\\
    &\approx
    -\frac{ e \bra{+} \mu_{\rm NH_3} \ket{-} \cos\theta_R }{ R^{2} }
    \int_{0}^{R}{\cal R}_{n' \mathrm{s}}(r){\cal R}_{n \mathrm{s}}(r) r^{2} \mathrm{d}r,
\end{align}
where $\theta_R = \cos^{-1}( \hat{\boldsymbol{\mu}}_{\rm NH_3} \cdot \hat{\boldsymbol{R}} )$ is the angle between the molecular dipole and collision axis, and ${\cal R}_{n \mathrm{s}}(r)$ is the Rydberg radial wavefunction obtained by numerical Numerov propagation.

\subsection{ Resonant collisional energy transfer }

The Hamiltonian during a scattering event, ignoring the center of mass terms as they constitute only a constant energy shift, is given by
\begin{align} \label{eq:collision_hamiltonian}
    H
    &=
    H_0 + V_{\rm CD}(\boldsymbol{r}, \boldsymbol{R}) \nonumber\\
    &=
    H_{\rm He}
    +
    H_{\rm NH_3}
    +
    \frac{ \boldsymbol{P}^2 }{ 2 \mu }
    +
    V_{\rm CD}(\boldsymbol{r}, \boldsymbol{R}),
\end{align}
where $\boldsymbol{P}$ is the relative momentum between the He$^+$ ion core and NH$_3$ molecule, and $\mu = m_{\rm He} m_{\rm NH_3} / ( m_{\rm He} + m_{\rm NH_3} )$ is the reduced mass of the collision partners.
Apart from energy, the collision must conserve parity of the total wavefunction in all relevant collisional degrees of freedom. 
By treating $R$ as an adiabatic coordinate, the combined He-NH$_3$ molecular state for Eq.~(\ref{eq:collision_hamiltonian}) can be expressed in the basis
\begin{align} \label{eq:state_basis}
    \ket{ n, \ell, m_{\ell} }\ket{ \Pi }\ket{ L, m_L },
\end{align}
where $\bra{ \hat{\boldsymbol{r}} }\ket{ L, m_L } = Y_{L, m_L}(\hat{\boldsymbol{R}})$ are the partial waves associated to the mechanical angular momentum of the collision. 
In principle, NH$_3$ prepared in the $J=K=1$ excited rotational state requires its parity doublet vibrational modes to be appropriately symmetrized along with the rotational and H$_3$ nuclear spin degrees of freedom \cite{Green80energy}. 
However, transitions between these latter states are far detuned in energy, allowing their treatment as spectator states so we suppress further explicit reference to them.  
For the near resonant $\ket{ 65\mathrm{s},- } \rightarrow \ket{ 66\mathrm{s}, + }$ transition, the NH$_3$ molecule undergoes an odd to even parity state, electric dipole transition, while a monopole transfer occurs between even parity Rydberg states of the He atom. 
Total parity must, therefore, be conserved through the collisional partial waves $\ket{ L, m_L }$, that can be coupled through the anisotropy of the interactions.

Of interest here is the transition between $\ket{ \mathrm{i} } = \ket{ 65\mathrm{s} }\ket{ - }$ and $\ket{ \mathrm{f} } = \ket{ 66\mathrm{s} }\ket{ + }$, that are close to resonant, but detuned by $\sim 40$ MHz.
Fortunately, this detuning is easily overcome by the collisional kinetic energy of the gas, which has a temperature of $\sim 100$ mK ($\equiv 2$ GHz).  
In contrast, the rotational splittings are $\gtrsim 100$ GHz in NH$_3$, making rotational state changing highly energetically suppressed. As such, we assume that the molecular dipole adiabatically orients itself to the local electrical field generated by the He$^+$ ion and the Rydberg electron as $R$ varies \cite{zhelyazkova21multipole}. In the parlance of molecular quantum mechanics, the initial molecular rotational state is gradually dressed by its neighboring states due to the electron/ion electric field upon approach, but never transitions to another dressed state so that it returns to its initial field-free state as it exits the collision.  
This approximation allows us to adopt a two-state model for NH$_3$ with only its lowest lying inversion doublet states.  
Even so, the local field-oriented molecular dipole still generally makes an angle with the collision coordinate $\theta_R = \cos^{-1}( \hat{\boldsymbol{\mu}}_{\rm NH_3} \cdot \hat{\boldsymbol{R}} )$, causing the potential to be anisotropic $V_{\rm CD}(\boldsymbol{R}) \sim \cos\theta_R$. 
This angular dependence induces the necessary dipole transitions for which $L' - L = \pm 1$, via the matrix elements:
\begin{align}
    \langle L', m'_L | \cos\theta_R | L, m_L \rangle
    &=
    (-1)^{m'_L}
    \sqrt{ (2 L' + 1) (2 L + 1) } \nonumber\\
    &\:\: \times 
    \begin{pmatrix}
        L' & 1 & L \\
        -m'_L & 0 & m_L 
    \end{pmatrix}
    \begin{pmatrix}
        L' & 1 & L \\
        0 & 0 & 0 
    \end{pmatrix},
\end{align}
in collisional partial waves of opposite parity, allowing the RET process to conserve total parity.

Nevertheless, we ignore the interaction anisotropy in our derivation by virtue of the RET process involving high partial wave scattering (further discussed below), allowing us to approximate $L' = L \pm 1 \approx L$. 
Moreover, a large number of $\ket{ L, m_L }$ states are typically involved in this regime, so we treat the scattering states directly in real space $\ket{ \nu }\ket{ \boldsymbol{K}_{\nu} } = \ket{ n, \ell, m_{\ell} }\ket{ \Pi }\ket{ \boldsymbol{K}_{\nu} }$ where $\ket{ \boldsymbol{K}_{\nu} }$ are relative momentum eigenstates with wavenumber $K_{\nu} = \sqrt{ 2\mu (E - \varepsilon_{\nu})/\hbar^2 }$, associated with the relative kinetic energy $E$ above threshold $\varepsilon_{\nu}$.

Scattering can be formulated in the interaction picture by first defining the potential $V_{\rm CD}(\boldsymbol{R}; t) = e^{i H_0 t / \hbar} V_{\rm CD}(\boldsymbol{R}) e^{-i H_0 t / \hbar}$, whereby the state-to-state overlap is then given to first-order by
\begin{align}
    & \bra{ \mathrm{f}; \boldsymbol{K}_{\mathrm{f}} }
    U(t-t_0)
    \ket{ \mathrm{i}; \boldsymbol{K}_{\mathrm{i}} } \nonumber\\
    =&\:
    \delta_{\mathrm{f}, \mathrm{i}}\,
    \delta^2( \boldsymbol{K}_{\mathrm{f}} - \boldsymbol{K}_{\mathrm{i}} ) 
    -
    \frac{ i }{ \hbar }
    \int_{t_0}^t 
    \mathrm{d}\tau\,
    \bra{ \mathrm{f}; \boldsymbol{K}_{\mathrm{f}} }
    V_{\rm CD}(\boldsymbol{R}; \tau)
    \ket{ \mathrm{i}; \boldsymbol{K}_{\mathrm{i}} } \nonumber\\
    =&
    -
    \frac{ i }{ \hbar }
    \int_{t_0}^t 
    \mathrm{d}\tau\,
    e^{ -i ( \omega_{|{\mathrm{i}; \boldsymbol{K}_{\mathrm{i}}}\rangle} - \omega_{|{\mathrm{f}; \boldsymbol{K}_{\mathrm{f}}}\rangle} ) \tau' }
    \bra{ \mathrm{f}; \boldsymbol{K}_{\mathrm{f}} }
    V_{\rm CD}(\boldsymbol{R})
    \ket{ \mathrm{i}; \boldsymbol{K}_{\mathrm{i}} },
\end{align}
where $\hbar \omega_{\ket{\nu; \boldsymbol{K}_{\nu}}} = { \hbar^2 K_{\nu}^2 / (2 \mu) } + \varepsilon_{\nu}$.
To safely apply the first-order Born approximation made above, it is important to determine its validity which we now do. Taking the range of the charge-dipole interaction to be the size of the Rydberg orbit $r_{\rm Ryd} \approx 2 n^2 a_0$, we expect that the scattered wave is only weakly perturbed by the interaction potential when
\begin{align}
    \left(
    \frac{ \hbar^2  }{ m\,r_{\rm Ryd}^2 }
    \right)^{-1}
    \max_{R} \bra{ \mathrm{f} } V_{\rm CD}(\boldsymbol{R}) \ket{ \mathrm{i} }
    &\ll
    k\,r_{\rm Ryd},
\end{align}
for which, taking the transition matrix element at $R = 100$ nm, we obtain a ratio of $\approx 3 \times 10^{-3} \ll 1$ for the left-hand side above, divided by the right-hand side. This lands us safely in the regime for a high energy approximation. 
The Born approximation has also previously been shown to be valid in Rydberg atoms with high $n$ colliding with other atoms \cite{Hickman1978theory, Hickman1979relation}. 

The scattering $S$-matrix is then computed as
\begin{align}
    S_{ \ket{ \mathrm{i}; \boldsymbol{K}_{\mathrm{i}} } \rightarrow \ket{ \mathrm{f}; \boldsymbol{K}_{\mathrm{f}} } }
    &=
    \underset{t \rightarrow \infty}{\lim_{t_0 \rightarrow -\infty}}
    \bra{ \mathrm{f}; \boldsymbol{K}_{\mathrm{f}} }
    U(t-t_0)
    \ket{ \mathrm{i}; \boldsymbol{K}_{\mathrm{i}} } \nonumber\\
    &=
    -\frac{ 2 \pi i }{ \hbar } 
    \delta( \omega_{|{\mathrm{f}; \boldsymbol{K}_{\mathrm{f}}}\rangle} - \omega_{|{\mathrm{i}; \boldsymbol{K}_{\mathrm{i}}}\rangle} ) \nonumber\\
    &\quad\quad\quad\quad \times
    \bra{ \mathrm{f}; \boldsymbol{K}_{\mathrm{f}} }
    V_{\rm CD}(\boldsymbol{R})
    \ket{ \mathrm{i}; \boldsymbol{K}_{\mathrm{i}} } \nonumber\\
    &=
    -2 \pi i \rho( E ) 
    \bra{ \mathrm{f}; \boldsymbol{K}_{\mathrm{f}} }
    V_{\rm CD}(\boldsymbol{R})
    \ket{ \mathrm{i}; \boldsymbol{K}_{\mathrm{i}} },
\end{align}
where $\rho(E)$ is the energy density of scattering states, which will be handled by choosing $\ket{ \boldsymbol{K} }$ to be energy normalized $\bra{ \boldsymbol{K}' }\ket{ \boldsymbol{K} } = \delta( E - E' ) \delta^2( \hat{\boldsymbol{K}} - \hat{\boldsymbol{K}}' )$, instead of unit normalized as assumed by the equations above.
As for the overlap integrals, we first perform a partial wave expansion of the energy normalized momentum eigenstates
\begin{align}
    \bra{ \boldsymbol{R} }\ket{ \boldsymbol{K} }
    &=
    \sqrt{ \frac{ 2 \mu }{ \pi \hbar^2 K } }
    e^{ i \boldsymbol{K} \cdot \boldsymbol{R} } \\
    &=
    \sqrt{ \frac{ 2 \mu }{ \pi \hbar^2 K } } 
    4 \pi \sum_{L, m_L}
    i^L 
    j_L( K R )
    Y_{L,m_L}(\hat{\boldsymbol{K}})
    Y_{L,m_L}^*(\hat{\boldsymbol{R}}), \nonumber
\end{align}
which implies the overlap integrals
\begin{widetext}
\begin{align}
    \bra{\nu'; \boldsymbol{K}_{\nu'} }
    V_{\rm CD}(\boldsymbol{R})
    \ket{ \nu; \boldsymbol{K}_{\nu} }
    &=
    \frac{ 2 \mu }{ \pi \hbar^2 \sqrt{ K_{\nu} K_{\nu'} } }
    16 \pi^2 
    \sum_{L', m'_L}
    \sum_{L, m_L}
    i^{L - L'}
    Y_{L',m'_L}^*(\hat{\boldsymbol{K}}') 
    Y_{L,m_L}(\hat{\boldsymbol{K}}) \nonumber\\
    &\quad\quad\quad\quad \times
    \int 
    Y_{L',m'_L}(\hat{\boldsymbol{R}})
    \cos\theta_R
    Y_{L,m_L}^*(\hat{\boldsymbol{R}})\,
    \mathrm{d}^2\hat{\boldsymbol{R}} \nonumber\\
    &\quad\quad\quad\quad \times
    \bra{\nu' }
    \int
    j_{L'}( K_{\nu'} R )
    V_{\rm CD}(R)
    j_L( K_{\nu} R )
    R^2\, \mathrm{d}R
    \ket{ \nu } \nonumber\\
    &\approx
    \frac{ 2 \mu }{ \pi \hbar^2 \sqrt{ K_{\nu} K_{\nu'} } }
    16 \pi^2 
    \sum_{L, m_L}
    Y_{L,m_L}^*(\hat{\boldsymbol{K}}') 
    Y_{L,m_L}(\hat{\boldsymbol{K}}) \nonumber\\
    &\quad\quad\quad\quad \times
    \bra{\nu' }
    \int
    j_{L}( K_{\nu'} R )
    V_{\rm CD}(R)
    j_L( K_{\nu} R )
    R^2\, \mathrm{d}R
    \ket{ \nu } \nonumber\\
    &\approx 
    \frac{ 32 \pi \mu }{ \hbar^2 \sqrt{ K_{\nu} K_{\nu'} } }
    \sum_{L, m_L}
    Y_{L,m_L}^*(\hat{\boldsymbol{K}}') 
    Y_{L,m_L}(\hat{\boldsymbol{K}}) \nonumber\\
    &\quad\quad\quad\quad \times 
    \bra{\nu' }
    \int
    \sin( K_{\nu'} R - L' \pi/2 )
    V_{\rm CD}(R)
    \sin( K_{\nu} R - L \pi/2 )\,
    \mathrm{d}R
    \ket{ \nu } \nonumber\\
    &=
    \frac{ 32 \pi \mu }{ \hbar^2 \sqrt{ K_{\nu} K_{\nu'} } }
    \sum_{L, m_L}
    Y_{L,m_L}^*(\hat{\boldsymbol{K}}') 
    Y_{L,m_L}(\hat{\boldsymbol{K}}) \nonumber\\
    &\quad\quad \times 
    \bra{\nu' }
    \frac{ 1 }{ 2 }
    \int
    \left[
    \cos( (K_{\nu'} - K_{\nu}) R )
    -
    \cos( (K_{\nu'} + K_{\nu}) R - L \pi )
    \right]
    V_{\rm CD}(R)\,
    \mathrm{d}R
    \ket{ \nu } \nonumber\\
    &\approx 
    \frac{ 32 \pi \mu }{ \hbar^2 \sqrt{ K_{\nu} K_{\nu'} } }
    \sum_{L, m_L}
    Y_{L,m_L}^*(\hat{\boldsymbol{K}}') 
    Y_{L,m_L}(\hat{\boldsymbol{K}}) 
    \bra{\nu' }
    \frac{ 1 }{ 2 }
    \int
    \cos\left[
    \left( K_{\nu'} - K_{\nu} \right) R
    \right]
    V_{\rm CD}(R)\,
    \mathrm{d}R
    \ket{ \nu },
\end{align}
\end{widetext}
having taken that $\cos[(K_{\nu'} + K_{\nu}) R - L \pi ]$ oscillates rapidly compared to $\cos[ (K_{\nu'} - K_{\nu}) R ]$, and so vanishes. 
This rotating wave approximation is valid so long as all other transitions in He are further detuned from the $\ket{-} \rightarrow \ket{+}$ interval than the $\ket{ 65\mathrm{s}} \rightarrow \ket{ 66\mathrm{s}}$ transition, which is true in the case of He. 
The wavelengths associated to the inelastic momentum transfer $2 \pi (K_{\mathrm{f}} - K_{\mathrm{i}})^{-1}$ would be comparable to the Rydberg orbit ($\sim$ hundreds of nanometers), making both terms in the integrand slowly varying over the interaction region and significant.   
For concreteness, we find that $2 \pi (K_{\mathrm{i}} - K_{\mathrm{f}})^{-1} \approx 270$~nm. 
On the other hand, the actual collision occurs at comparatively large momentum $K_{\mathrm{i}}$, making a natural length scale over which the collision takes place to be $2 \pi K_{\mathrm{i}}^{-1} \approx 3.5$~nm, comparable to the Rydberg electron de Broglie wavelength.

Exploiting the approximate spherical symmetry of the problem, we identify $\sum_{L, m_L} Y_{L,m_L}(\hat{\boldsymbol{K}}') Y_{L,m_L}(\hat{\boldsymbol{K}}) = \delta^2(\hat{\boldsymbol{K}} - \hat{\boldsymbol{K}}')$,
where one might consider replacing $\delta^2(\hat{\boldsymbol{K}} - \hat{\boldsymbol{K}}')$ with an angular ``density" $(4 \pi)^{-1}$. 
However, the anisotropy and change in kinetic energy from the inelastic transition is likely associated with a small but non-negligible scattering angle $\alpha$. 
To estimate this angle, we consider that the detuning $\Delta_{\mathrm{i},\mathrm{f}} = \varepsilon_{\mathrm{f}} - \varepsilon_{\mathrm{i}}$ between the initial and final states induces a momentum change $\delta\boldsymbol{K} = \hat{\delta\boldsymbol{K}} \sqrt{ 2 \mu \Delta_{\mathrm{i},\mathrm{f}} / \hbar^2}$ during the collision so that $\boldsymbol{K}_{\mathrm{f}} = \boldsymbol{K}_{\mathrm{i}} - \delta\boldsymbol{K}$, from which we can derive the relation: 
\begin{align}
    \Delta_{\mathrm{i},\mathrm{f}}
    &=
    \frac{ \hbar^2 K_{\mathrm{i}}^2 }{ 2\mu }
    -
    \frac{ \hbar^2 K_{\mathrm{f}}^2 }{ 2\mu } \nonumber\\
    &=
    \frac{ \hbar^2 K_{\mathrm{i}}^2 }{ 2\mu }
    -
    \frac{ \hbar^2 (\boldsymbol{K}_{\mathrm{i}} - \delta\boldsymbol{K})^2 }{ 2\mu } \nonumber\\
    &=
    \frac{ \hbar^2 }{\mu }
    K_{\mathrm{i}}\, \delta{K}
    \cos\alpha
    -
    \frac{ \hbar^2\, \delta{K}^2 }{ 2\mu } \nonumber\\
    &=
    \frac{ \hbar^2 }{\mu }
    K_{i} \sqrt{ 2 \mu \Delta_{\mathrm{i},\mathrm{f}} / \hbar^2 }
    \cos\alpha
    -
    \Delta_{\mathrm{i},\mathrm{f}}, \nonumber\\
    \Rightarrow\quad 
    \cos\alpha
    &=
    \sqrt{ \frac{ 2 \mu \Delta_{\mathrm{i},\mathrm{f}} }{ \hbar^2 K_{\mathrm{i}}^2 } }.
\end{align}
Therefore, instead of assuming perfectly collinear incident and outgoing momenta, we take that $\hat{\boldsymbol{K}}'$ deviates from $\hat{\boldsymbol{K}}$ by $\alpha$ which subtends the solid angle $\delta\Omega_{\alpha} = \int_{0}^{2\pi} \int_{0}^{\alpha} \sin\theta_K\, \mathrm{d}\theta_K\, \mathrm{d}\phi_K$, providing us with the natural replacement:
\begin{align}
    \delta^2(\hat{\boldsymbol{K}} - \hat{\boldsymbol{K}}')
    \quad\rightarrow\quad
    \frac{ \delta\Omega_{\alpha} }{ 4\pi }
    &=
    \frac{ 1 }{ 2 }    
    \left(
    1 - \sqrt{ \frac{ 2 \mu \Delta_{\mathrm{i},\mathrm{f}} }{ \hbar^2 K_{\mathrm{i}}^2 } }
    \right),
\end{align}
and the matrix element
\begin{align}
    & \bra{\nu'; K_{\nu'} }
    V_{\rm CD}(R)
    \ket{ \nu; K_{\nu} }
    \approx
    \frac{ 8 \pi \mu }{ \hbar^2 \sqrt{ K_{\nu} K_{\nu'} } }
    \left(
    1 - \sqrt{ \frac{ 2 \mu \Delta_{\mathrm{i},\mathrm{f}} }{ \hbar^2 K_{\mathrm{i}}^2 } }
    \right) \nonumber\\
    &\quad\quad \times 
    \int
    \bra{\nu' }
    V_{\rm CD}(R)
    \ket{ \nu }
    \cos\left[
    \left( K_{\nu'} - K_{\nu} \right) R
    \right]\,
    \mathrm{d}R,
\end{align}
where we drop any further reference to $\hat{\boldsymbol{K}}$.  
As a result, the $S$-matrix element can be approximated as
\begin{align} \label{eq:S_matrix_element}
    & S_{ |{ \mathrm{i}; \boldsymbol{K}_{\mathrm{i}} }\rangle \rightarrow |{ \mathrm{f}; \boldsymbol{K}_{\mathrm{f}} }\rangle }
    \approx
    -2 \pi i
    \frac{ 8 \pi \mu }{ \hbar^2 \sqrt{ K_{\mathrm{i}} K_{\mathrm{f}} } }
    \left(
    1 - \sqrt{ \frac{ 2 \mu \Delta_{\mathrm{i},\mathrm{f}} }{ \hbar^2 K_{\mathrm{i}}^2 } }
    \right) \nonumber\\
    &\quad\:\: \times
    \int_{R_{\min}}^{R_{\max}}
    \bra{ \mathrm{f} }
    V_{\rm CD}(R)
    \ket{ \mathrm{i} }
    \cos\left[
    \left( K_{\mathrm{f}} - K_{\mathrm{i}} \right) R
    \right]\,
    \mathrm{d}R.
\end{align}
As indicated above, we integrate over $R$ starting from $R_{\min}$ out to $R_{\max} = 1000$~nm where the transition matrix element is negligibly small.  
We determine $R_{\min}$ in terms of a minimum impact parameter $b$ below which we expect chemistry to occur in a classical capture model \cite{tsikritea22capture}. 
At short-range, the ammonia molecule predominantly experiences the field generated by the He$^+$ ion, so that the centrifugal energy in the collision is sufficient to overcome the ion-dipole attraction when:
\begin{align}
    \frac{ {\cal L}^2 }{ 2 \hbar^2 \mu R^2 }
    &>
    \frac{ \mu_{\rm NH_3} e }{ 4 \pi \epsilon_0 R^2 }, \\
    \Rightarrow\quad 
    {\cal L}^2
    &>
    \frac{ \mu \hbar^2 \mu_{\rm NH_3} e }{ 2 \pi \epsilon_0 },
\end{align}
where ${\cal L}$ is the classical angular momentum. Utilizing the semiclassical approximation, that ${\cal L} = \hbar ( L + 1/2 )$, we estimate that the $L \gtrsim 210$ partial wave is needed to evade chemistry. 
In the the parlance of quantum scattering, the addition of $\hbar^2 L (L + 1) / (2 \mu R^2)$ with $L = 210$ to the adiabatic curve that asymptotically connects to $\ket{65\mathrm{s}, +}$, is barriered against the $\sim 1.5$ GHz collision energy. 
Identifying the collisional partial wave with an impact parameter $L = K b$, the minimum impact parameter at a temperature of $\approx 70$~mK is approximately $b \gtrsim 120$~nm. 
In practice, we utilize $R_{\min} = 100$~nm, where given semiclassical scattering with an almost negligible potential, an NH$_3$ molecule has about a $1 \%$ probability of entering the classical capture sphere of radius $R=100$~nm over any other point inside the Rydberg orbit of $\approx 450$~nm. 
This low probability renders chemical processes below the ionization detection limit.
The matrix element $\bra{ \mathrm{i} } V_{\rm CD}(R) \ket{ \mathrm{f} }$ is plotted as a function of $R$ in Fig.~\ref{fig:theory_figure}, for both $\ket{ \mathrm{f} } = \ket{ 66\mathrm{s}, + }$ (dark blue) and $\ket{ 64\mathrm{s}, - }$ (light red). 
The integral cross section is then obtained from
\begin{align}
    \sigma_{ \ket{ \mathrm{i} } \rightarrow \ket{ \mathrm{f} } }
    &=
    \frac{ \pi }{ K_{\mathrm{i}}^2 }
    \abs{ 
    T_{ |{ \mathrm{i}; K_{\mathrm{i}} }\rangle \rightarrow |{ \mathrm{f}; K_{\mathrm{f}} }\rangle } }^2. 
\end{align}
where $T_{ |{ \mathrm{i}; K_{\mathrm{i}} }\rangle \rightarrow |{ \mathrm{f}; K_{\mathrm{f}} }\rangle } = \delta_{\mathrm{i}, \mathrm{f}}
    -
    S_{ |{ \mathrm{i}; K_{\mathrm{i}} }\rangle \rightarrow |{ \mathrm{f}; K_{\mathrm{f}} }\rangle }$ is the $T$-matrix.

From this cross section, we can compute an expected rate for the $\ket{ 65\mathrm{s},- } \rightarrow \ket{ 66\mathrm{s}, + }$ transition in a thermal sample of NH$_3$ per He atom, via the formula
\begin{subequations}
\begin{align}
    \Gamma_{ \ket{ \mathrm{i} } \rightarrow \ket{ \mathrm{f} } }
    &=
    n_{\rm NH_3}
    \beta_{ \ket{ \mathrm{i} } \rightarrow \ket{ \mathrm{f} } }, \\
    \beta_{ \ket{ \mathrm{i} } \rightarrow \ket{ \mathrm{f} } }
    &=
    \int
    \sigma_{ \ket{ \mathrm{i} } \rightarrow \ket{ \mathrm{f} } }(K)
    \frac{ \hbar K }{ \mu }
    c(\boldsymbol{K})\,
    \mathrm{d}^3\boldsymbol{K},
\end{align}
\end{subequations}
where $n_{\rm NH_3}$ is the local density of NH$_3$ around each He atom and $c(\boldsymbol{K})$ is the distribution of relative momenta between the NH$_3$ molecules and He.  
Assuming that there are many more NH$_3$ molecules than He Rydberg atoms, which is the case in the experiments, we expect that the probability of a He Rydberg atom remaining in the $\ket{ 65\mathrm{s} }$ state is given by the rate equation:
\begin{subequations}
\begin{align}
    \frac{ \mathrm{d}\mathbb{P}_{\rm 65\mathrm{s}} }{ \mathrm{d}t }
    &\approx
    -n_{\rm NH_3}
    \beta_{ \ket{ \mathrm{i} } \rightarrow \ket{ \mathrm{f} } }
    \mathbb{P}_{\rm 65\mathrm{s}}, \\
    \Rightarrow\quad 
    \mathbb{P}_{\rm 65\mathrm{s}}(t)
    &\approx
    \exp({ 
    -n_{\rm NH_3}
    \beta_{ |{ \mathrm{i} }\rangle \rightarrow |{ \mathrm{f} }\rangle } t 
    }).
\end{align}
\end{subequations}
Detection by electric field ionization is performed after an interaction time $\Delta t = 12$ $\mu$s, before which collisions are allowed to occur. The expected probability that any given Helium atom transitions to the $\ket{ 66\mathrm{s} }$ state is thus given by
\begin{align}
    \mathbb{P}_{ |{ 66\mathrm{s} }\rangle }
    &\approx 
    1
    -
    \exp\left(
    -n_{\rm NH_3}
    \beta_{ |{ \mathrm{i} }\rangle \rightarrow |{ \mathrm{f} }\rangle } 
    \Delta{t}
    \right).
\end{align}
With the local NH$_3$ density in the experiments taken to be $n_{\rm NH_3} = (1.5 \pm 0.4) \times 10^{10}$ cm$^{-3}$, and a mean collision speed of $\langle v \rangle = 19.3 \pm 2.6$ m/s,  we obtain $\mathbb{P}_{ |{ 66\mathrm{s} }\rangle } \approx 17 \pm 4 \%$.

Similarly, we estimate the probability of a transition to the $\ket{ 64\mathrm{s} }$ state to be $\mathbb{P}_{ |{ 64\mathrm{s} }\rangle } \approx 0.12 \pm 0.03 \%$, which is strongly suppressed due to its much larger detuning of $\varepsilon_{| 65\mathrm{s},+ \rangle} - \varepsilon_{|64\mathrm{s},-\rangle} \approx 1.166$~GHz.  
This suppression can be intuited from the integral in Eq.~(\ref{eq:S_matrix_element}), showing that for large detuning, the cosine term oscillates rapidly and averages over the slow-varying features of the transition matrix element.

\end{document}